\documentclass[12pt]{article}
\usepackage{amsmath,amsfonts,amssymb,amsthm,latexsym}
\usepackage{epsfig}
\usepackage{graphicx}
\usepackage{subfigure}
\usepackage{mathabx}
\usepackage{bbm}

\newcommand{\ds}{\displaystyle}

\newcommand{\ket}[1]{|\kern.3ex#1\kern.3ex\rangle}
\newcommand{\bra}[1]{\langle\kern.3ex #1 \kern.3ex|}
\newcommand{\scalar}[2]{\langle\kern.3ex #1 \kern.3ex|\kern.3ex#2\kern.3ex\rangle}

\newcommand{\ii}{\mathsf{i}}

\def\N{\mathbb{N}}
\def\C{\mathbb{C}}

\def\lg{\langle }
\def\rg{\rangle }

\def\adg{a^{\dag}}

\def\vk{\varkappa}

\def\ud{\mathrm{d}}

\def\sfP{\mathsf{P}}

\def\hN{\hat N}
\def\adgh{\left(a^h\right)^{\dag}}
\def\bsh{\boldsymbol{\mathsf{h}}}
\def\adgh{\left(a^{\bsh}\right)^{\dag}}

\def\bu{\mathbbm{1}}

\numberwithin{equation}{section}

\begin{document}

%%%%%%%%%%%%%%%%%%%%%%%%%%%%%%%%%%%

\thispagestyle{empty}
\hfill \today

\vspace{2.5cm}

\begin{center}
\bf{\LARGE
SU$(1,1)$-displaced coherent states, photon counting and squeezing
}
\end{center}

\bigskip\bigskip

\begin{center}
J.-P. Gazeau$^{1,2}$, M.A. del Olmo$^2$.
\end{center}

\begin{center}
$^1${\sl Universit\'e Paris Cit\'e, CNRS, Astroparticule et Cosmologie, F-75013 Paris, France}\\
\medskip

$^2${\sl Departamento de F\'{\i}sica Te\'orica and IMUVA, Universidad de
Valladolid, \\
E-47011, Valladolid, Spain.}\\
\medskip

{e-mail: gazeau@apc.in2p3.fr\,,
\; marianoantonio.olmo@uva.es}

\end{center}

\bigskip

\bigskip

\begin{abstract}
We revisit the Perelomov SU$(1,1)$ displaced coherent states states as possible quantum states of  light. We disclose interesting statistical aspects of these states in relation with photon counting and squeezing. In the non-displaced case we discuss the efficiency of the photodetector as inversely proportional to the parameter $\varkappa$ of the discrete series of unitary irreducible representations of SU$(1,1)$.  In the displaced case, we study the counting and squeezing properties of the states in terms of $\varkappa$ and the number of photons in the original displaced state.  We finally examine the quantization of a classical radiation field which  is based on these families of coherent states. The procedure yields  displacement operators  which might allow to prepare such states  in the way proposed by Glauber for the standard coherent states.
\end{abstract}

%%%%%%%%%%%%%%%%%%%%%%%%%%%%%%%%%%%%%%%%%%%%%%%%%%%%%%
%%%%%%%%%%%%%%%%%%%%%%%%%%%%%%%%%%%%%%%%%%%%%%%%%%%%%%
\section{Introduction}
\label{intro}

Over the last decades,  with the works by W\'odkiewicz and Eberly \cite{wodeb85}, Gerry \cite{gerry85,gerry87}, and many other authors, see reviews like \cite{dellanno06,gazeaurev18},  Perelomov SU$(1,1)$ coherent states (CS) \cite{perel72,perel86,gazeaubook09,gazolmo18-1} have been considered from a quantum optics point of view. In this paper we examine the   SU$(1,1)$ Perelomov displaced-coherent states  from the quantum optics perspective and study the rôle of two parameters. The first one  labels the discrete series of the unitary irreducible representations (UIR) of SU$(1,1)$ and beyond, and the second one is the number of photons in the fiducial state used to build the CS's. A large part of the mathematical background used in this paper is already present in the well known Perelomov monograph \cite{perel86} and previous publications by the authors \cite{gazeaurev18,gazolmo18-1,curadoetal20}.

In Section \ref{genset} we introduce a specific class of normalized non-standard coherent states  resolving the identity in the Hilbert-Fock space  of one-mode photons, and we give a short account of statistical properties which will be studied in the next sections. 
In Section \ref{SUIICS} we adapt the Perelomov  SU$(1,1)$-CS's to the quantum optics framework and we study  their statistical properties  and squeezing properties with regard to the variation of  two parameters, $\varkappa$ and $s$, the former labelling the considered UIR's of SU$(1,1)$, and the latter one being the number of photons present in the original state used to build the SU$(1,1)$-CS's. 
  The content of Section \ref{ANCSQsection} concerns the possible role of these  types of  CS  in the quantization of classical solutions of the Maxwell equations and the corresponding quadrature portraits. 
Some promising features of this CS quantization are discussed in Section \ref{conclu}.  

%%%%%%%%%%%%%%%%%%%%%%%%%%%%%%%%%%%%%%%%%%%%%%%%%%%%%%
%%%%%%%%%%%%%%%%%%%%%%%%%%%%%%%%%%%%%%%%%%%%%%%%%%%%%%

\section{General setting for coherent states in a wide sense}
\label{genset}
Before considering the SU$(1,1)$ case in the next sections, let us give a brief account of a recent review devoted to generalised or non-standard coherent states in quantum optics \cite{gazeaurev18}.   In the quantum modelling  of a monochromatic radiation,   the  kets $|n\rg$, where  $n=0, 1, 2,\dotsc,$ stands for the number of photons,  form an orthonormal basis of the Fock-Hilbert space  $\mathcal{H}$. 
For  such a basis, the familiar  \emph{annihilation}  operator  $a$ and its adjoint $a^{\dag}$, the  \emph{creation} operator, are defined by
\begin{equation}
\label{acaadag}
a | n \rg = \sqrt{n} |n -1\rg\, , \quad \adg |n \rg = \sqrt{n+1} |n +
1\rg\, , 
\end{equation}
together with the action of $a$ on the vacuum state given by 
$ a |0 \rg = 0$. They obey the so-called canonical commutation rule, i.e.,   $[a,\adg]= \bu$.
In this context, the photon number operator $\hN = \adg a $ is diagonal in the  basis $\{|n\rg, \ n \in \N\}$, with spectrum $\N$: $\hN |n\rg = n | n \rg$.

Many of the generalisations of the standard Glauber-Sudarshan-CS (GSCS)  \cite{glauber63-1,glauber63-3,sudarshan63} encountered in quantum optics or in quantum mechanics are  one-mode Fock states  of the  form  	
\begin{equation}
\label{anclass}
\begin{array}{lll}
|\alpha;\bsh;s\rg &=&\ds\sum_{n=0}^{s} {\bar\alpha}^{s-n}\,h_n(\vert\alpha\vert^2) |n\rg\\[0.2cm] 
&&\qquad
\ds + \sum_{n=s+1}^{\infty} \alpha^{n-s}\,h_n(\vert\alpha\vert^2) |n\rg\\[0.4cm] \ds
&\equiv &\ds \sum_{n=0}^{\infty} \phi_n(\alpha,\bar\alpha) |n\rg \, , 
\end{array}\end{equation}
where  $s\in \N$,  
$\alpha\in\C$  lies in the centred open disk $\mathcal{D}_R$ of radius $0<R\leq \infty$  (i.e., $\vert \alpha\vert ^2<  R^2$) and $\bar\alpha$ is the complex conjugate of $\alpha$. In the sequel we will use the notation $u= \vert \alpha\vert^2$. The sequence $\bsh:= \left( h_n\right)\, , \, n=0,1,2,\dotsc$,  of real-valued functions 
\begin{equation}
\label{hnu}
[0,R^2) \ni u  \mapsto h_n(u)
\end{equation}
is requested to obey the following three  conditions:

\noindent 
1.- {\it Sum normalisation}:
\begin{equation}
\label{condnorm}
\ds\sum_{n=0}^{\infty} \vert \phi_n(\alpha, \bar\alpha)\vert^2 \equiv \ds \sum_{n=0}^{s} u^{s-n}\,\left(h_n(u)\right)^2 +   \sum_{n=s+1}^{\infty} u^{n-s}\,\left(h_n(u)\right)^2\\
=1\, . 
 \end{equation}
 
\noindent 
2.- {\it The function $\bar n_{\bsh} (u)$  defined by:
 \begin{equation}
 \label{onetoone} [0,R^2) \ni u \mapsto \bar n_{\bsh} (u):= \sum_{n=0}^{\infty} n\,\vert \phi(\alpha, \bar\alpha)\vert^2\,.
 \end{equation}
 is strictly increasing.}

\noindent 
3.- {\it Integral normalisation}:  there exists  a weight function $w^{\bsh}(u)$ on $[0,R^2)$ such that:
\begin{equation}\label{condortho}   \int_{0}^{R^2}\ud u\,w^{\bsh}(u)\, \vert \phi_n(\alpha, \bar\alpha)\vert^2 =1 \,, \quad n=0,1,2,\dotsc\, \,.
\end{equation}

Note that the GSCS's, i.e., the standard ones, are obtained by putting  in \eqref{anclass}:
\begin{equation}
\label{gscspc}
s=0\,, \qquad h_n(u)= \frac{e^{-u/2}}{\sqrt{n!}}\, , \qquad w^{\bsh}(u)=1\,.
\end{equation}
Then  $R=\infty$.

Condition  \eqref{condnorm} means that the vectors in \eqref{anclass} are normalised states in the Fock-Hilbert space spanned by the number states. It allows to interpret the discrete map  
\begin{equation}
n \mapsto   \sfP^{\bsh}_n(u)=\vert \phi_n(\alpha, \bar\alpha)\vert^2\, , 
\end{equation} 
with
\begin{equation}
\label{probdet}
\sfP^{\bsh}_n(u)
=\left\lbrace\begin{array}{ll}
   u^{s-n}\,\left(h_n(u)\right)^2   \;&  n=0,1,\dotsc\,, s  \\[0.4cm]
    u^{n-s}\,\left(h_n(u)\right)^2  \;&   n=s+1,\dotsc
\end{array}\, , \right.
\end{equation}
as a probability distribution on $\N$ with parameter $u$. It is precisely the probability of registering $n$ photons with a measuring device having maximal efficiency when the  light beam is in the non-standard  coherent state $|\alpha;\bsh;s\rg$.

Condition  \eqref{onetoone} expresses that  the expected value 
\begin{equation}
\label{avernb}
\lg\alpha;\bsh;s| \hat N |\alpha;\bsh;s\rg = \sum_{n=0}^{+\infty} n \,\sfP^{\bsh}_n(u)\equiv \bar n_{\bsh} (u) 
\end{equation}
of the number operator  is one-to-one function of $u$ and so can be inverted. Note that $\bar n = u$ in the  GSCS case  \eqref{gscspc}.

Condition  \eqref{condortho} implies the resolution of the identity in the Fock space, i.e.:
\begin{equation}\label{resid}
 \int_{\mathcal{D}_R}\frac{\ud^2\alpha}{\pi}\,w\left(\vert\alpha\vert^2\right)\, |\alpha;\bsh;s\rg\lg\alpha;\bsh;s| = \bu\, .
\end{equation}
This property holds because of the orthonormality relations 
\begin{equation}
\label{orthophi}
 \int_{\mathcal{D}_R}\frac{\ud^2\alpha}{\pi}\,w\left(\vert\alpha\vert^2\right)\, \overline{\phi_n(\alpha, \bar\alpha)}\,\phi_{n^{\prime}}(\alpha, \bar\alpha)= \delta_{n n^{\prime}}\,. 
\end{equation}
They are easily derived  from Fourier angular integration on the argument of $\alpha$ and  the  kind of moment problem solved by \eqref{condortho}.
The latter  allows us to interpret the map 
\begin{equation}\label{alprob}
\alpha \mapsto  \sfP^{\bsh}_n\left(\vert\alpha\vert^2\right)
\end{equation}
as an isotropic  probability distribution, with parameter $n$,  on the disk $\mathcal{D}_R$, equipped with the  measure $w\left(\vert\alpha\vert^2\right)\,\dfrac{\ud^2\alpha}{\pi}$. Equivalently, the map
\begin{equation}
\label{uprob}
[0,R^2) \ni u \mapsto  \sfP^{\bsh}_n(u)
\end{equation}
is a probability distribution on the interval $[0,R^2)$ equipped with the measure $w(u)\,\ud u$.

 The expectation value \eqref{avernb} of the number operator 
 can be viewed as the intensity (or energy up to a physical factor like $\hbar \omega$) of the state $|\alpha;\bsh;s\rg$ of the  quantum monochromatic radiation under consideration. An optical phase space associated with this ``non-standard-CS radiation'' may be defined as  the image of the map 
 \begin{equation}
\label{mapaln}
\mathcal{D}_R \ni \alpha \mapsto \xi_{\alpha}=\sqrt{\bar n_{\bsh}\left(\vert\alpha\vert^2\right)}\,e^{\ii \arg \alpha}\in \C\, .
\end{equation}

The fluctuations of the number of photons  about its mean value 
are quantified in terms of the standard deviation 
\begin{equation}
\Delta n_{\bsh}(u)=\sqrt{\overline{n^2_{\bsh}}-\left(\bar{n}_{\bsh} \right)^2}\, , \qquad 
\overline{n^2_{\bsh}}(u)=\sum_n n^2 \,\sfP^{\bsh}_n(u)\,.
\end{equation}

The distribution $n \mapsto  \sfP^{\bsh}_n(u)$ is then classified as: 
\begin{enumerate}
\item
Sub-poissonian for $\Delta n_{\bsh}<\sqrt{\bar{n}_{\bsh}}$, 

\item 
Poissonian for $\Delta n_{\bsh}=\sqrt{\bar{n}_{\bsh}}$ (e.g. standard case \eqref{gscspc})\,,   

\item 
Super-poissonian for $\Delta n_{\bsh}>\sqrt{\bar{n}_{\bsh}}$.
\end{enumerate}

The deviation of $\sfP^{\bsh}_n(u)$ from the Poisson distribution can be measured with the \textit{Mandel parameter} $Q^{\bsh}$ \cite{mandel_wolf70,mandel_wolf95}. The latter is defined by
\begin{equation}\label{mandel}
Q^{\bsh}:=\frac{\overline{n^2_{\bsh}}-\left(\bar{n}_{\bsh} \right)^2}{\bar{n}_{\bsh}} -1\,.
\end{equation}
At a given $u$  the distribution $\sfP^{\bsh}_n(u)$ is   sub-Poissonian if $Q^{\bsh}(u)<0$, 
Poissonian if $Q^{\bsh}(u)=0$, and super-Poissonian if $Q^{\bsh}(u)>0$.

%%%%%%%%%%%%%%%%%%%%%%%%%%%%%%%%%%%%%%%%%%%%%%%%%%%%%%
%%%%%%%%%%%%%%%%%%%%%%%%%%%%%%%%%%%%%%%%%%%%%%%%%%%%%%

\section{Perelomov SU$(1,1)$-displaced coherent states as optical CS}
\label{SUIICS}
Perelomov SU$(1,1)$ coherent states  are obtained through a SU$(1,1)$ unitary action on the vacuum, and more generally on a number state. The Fock-Hilbert space $\mathcal{H}$ is infinite-dimensional while the complex number $\alpha$ is restricted to lie in the open unit disk $\mathcal{D}:= \{\alpha  \in \C\, , \, \vert \alpha \vert <  R=1\}$. 
For $\varkappa> 1/2$ and $s\in \N$ we  define   the ``SU$(1,1)$-displaced  $s$-th coherent states''  (which should not be confused with the so-called photon-added CS, see for instance \cite{iqbal22} and references therein) as:
\begin{equation}
\label{morecssu11}
|\alpha; \varkappa; s\rg=  U^{\varkappa}(p( \bar \alpha))|s\rg= \sum_{n = 0}^{\infty} U^{\varkappa}_{ns}(p( \bar \alpha)) |n\rg\, ,
\end{equation}
where the $U^{\varkappa}_{ns}(p( \bar \alpha))$'s are matrix elements of the  UIR $U^{\varkappa}$ of SU$(1,1)$ in its discrete series and $p( \bar\alpha)$ is the particular 
matrix 
\begin{equation}
\label{palapha}
\begin{pmatrix}
  \left(1-\vert \alpha\vert^2\right)^{-1/2}    &   \left(1-\vert \alpha\vert^2\right)^{-1/2}\,\bar\alpha   \\
   \left(1-\vert \alpha\vert^2\right)^{-1/2} \,\alpha    &   \left(1-\vert \alpha\vert^2\right)^{-1/2} 
\end{pmatrix}\in \mathrm{SU}(1,1)\,. 
\end{equation}
They are given in terms of the Jacobi polynomials $P^{(\alpha , \beta)}_{n}\left( x \right)$ \cite{magnus66} as
\begin{align}
\label{Upz}
\nonumber   U^{\varkappa}_{ns}(p(\bar \alpha)) & = \left( \frac{n_<!\, \Gamma(2 \varkappa + n_>) }{n_>!\, \Gamma(2 \varkappa + n_<)} \right)^{1/2}  \left(1-\vert \alpha \vert^2\right)^{\varkappa} \, (\mathrm{sgn}(n-s))^{n-s} \\
&\qquad \times   P^{(n_> - n_<\, ,\, 2 \varkappa -1)}_{n_<}\left( 1-2\vert \alpha\vert^2 \right)\,\nonumber\\
&\qquad\qquad \times\left\lbrace\begin{array}{cc}
\alpha^{n-s}      & \mathrm{if}\ n_{>} =n  \\
 \bar\alpha^{s-n}     &   \mathrm{if}\ n_{>} =s
\end{array}\right.
\end{align}
with  $n_{\substack{
>\\
<}}  = \left\lbrace \begin{array}{c}
    \max      \\
       \min
\end{array}\right.\, (n,s) \geq 0$.

\noindent 
Hence, in the present case, the sequence of functions $\bsh:= \left( h_n\right)$ \eqref{hnu} is given by
\begin{equation}\label{SUhn}\begin{array}{lll}
\hskip-0.45cm h_n(u)= (1-u)^{\varkappa} \, (\mathrm{sgn}(n-s))^{n-s}\\ 
\hskip-0.45cm\;\;\times\left\lbrace\begin{array}{ll}
C(\varkappa,s,n) \,P^{(s - n\, ,\, 2 \varkappa -1)}_{n}( 1-2u)    &   \mathrm{if} \;  0\leq n \leq s \\[0.3cm]
 C^{-1}(\varkappa,s,n)  \,P^{(n - s\, ,\, 2 \varkappa -1)}_{s}( 1-2u)   & \mathrm{if}\; n \geq s+1
\end{array}
\right.
\end{array}\end{equation}
where $C(\varkappa,s,n) =\left( \dfrac{n!\, \Gamma(2 \varkappa + s) }{s!\, \Gamma(2 \varkappa + n)} \right)^{1/2}$.
The states $|\alpha; \varkappa; s\rg$ \eqref{morecssu11} solve the identity 
\begin{equation}
\label{runitmorecssu11}
\frac{2\varkappa-1}{\pi}\int_{\mathcal{D}}\frac{\ud^2 \alpha}{\left(1-\vert\alpha\vert^2\right)^2} \, |\alpha; \varkappa;s \rg \lg \alpha; \varkappa;s | = \bu\,.
\end{equation}
This follows  from \eqref{resid}, with  $R=1$ and     
\begin{equation}
\label{wsu11}
w(u)= \frac{2\varkappa-1}{(1-u)^2}\,, 
\end{equation}
which makes clear the prior restriction $\varkappa > 1/2$.  
Let us first consider the simplest case s=0 and, next,  the general  cases for  $s>0$.

\subsection{\underline {$s=0$}}
The corresponding CS's are well known \cite{perel72}. They read as:
\begin{equation}
\label{cssu11}
|\alpha;\varkappa;0\rg\equiv |\alpha;\varkappa\rg= \sum_{n=0}^{\infty}\alpha^n\,h_{n;\varkappa}\left(\vert\alpha\vert^2\right)\,|n\rg\, , 
\end{equation}
with $\quad h_{n;\varkappa}(u):=  \sqrt{\binom{2\varkappa -1 +n}{n}}\, (1-u)^{\varkappa}$. Note that we can rewrite  the  expression \eqref{cssu11}
in the form of a ``nonlinear'' CS:
\begin{equation}
\label{nlcssu11}
|\alpha;\varkappa\rg= (1-u)^{\varkappa} \sum_{n=0}^{\infty}\frac{\alpha^n}{\sqrt{x_n!}}\,|n\rg\, , \quad x_n= \frac{n}{2\varkappa -1 +n}\, . 
\end{equation}
The corresponding  detection probability distribution is negative binomial,  
\begin{equation}
\label{negbinomialdist}
n\mapsto \sfP^{(\varkappa ,0)}_n(u)= (1-u)^{2\varkappa}\,\binom{2\varkappa -1 +n}{n}\,u^n\,. 
\end{equation}
The average value of the number operator reads as
\begin{equation}
\label{avernphperel}
\bar n^{(\varkappa ,0)}(u)\equiv \bar n= 2\varkappa\,\frac{u}{1-u}\ \Leftrightarrow\ u= \frac{\bar n/2\vk}{1+\bar n/2\vk} \,. 
\end{equation}
The Mandel parameter \eqref{mandel} assumes its   simplest form, which is nonnegative and  independent of $\varkappa$:
\begin{equation}
\label{mandelsu11}
Q^{(\varkappa ,0)}(u)= \frac{u}{1-u}\, .
\end{equation}
Hence, the Perelomov SU$(1,1)$ CS for $s=0$ are super-Poissonian for all $u\in (0,1)$. This classical feature finds its explanation in  their ``thermal'' nature.   
Indeed, by introducing the ``efficiency'' $\eta :=\frac{1}{2 \vk}\in (0,1)$, the probability $\sfP^{(\varkappa ,0)}_n(u)$ in \eqref{negbinomialdist} is expressed in terms of the corrected average value $\bar N:=\eta\,\bar n = Q(u)$ as 
 \begin{equation}
\label{negbinomialdist1}
\widetilde{\sfP}^{(\varkappa ,0)}_n(\bar N)= (1+\bar N)^{-1/\eta}\binom{1/\eta-1 +n}{n}\, \left(\frac{\bar N}{1+\bar N}\right)^n\, .
\end{equation}
%%%%%%%%%%%%%%%%%%%%%
%%%%%%%%%%%%%%%%%%%%%
\begin{figure}[ht]
\centering
\subfigure[ $\sfP_n^{(1 ,0)}(u)$  \hskip 5cm(b) $\widetilde{\sfP}^{(1 ,0)}(\bar N)$]{\includegraphics[width=0.45\textwidth]{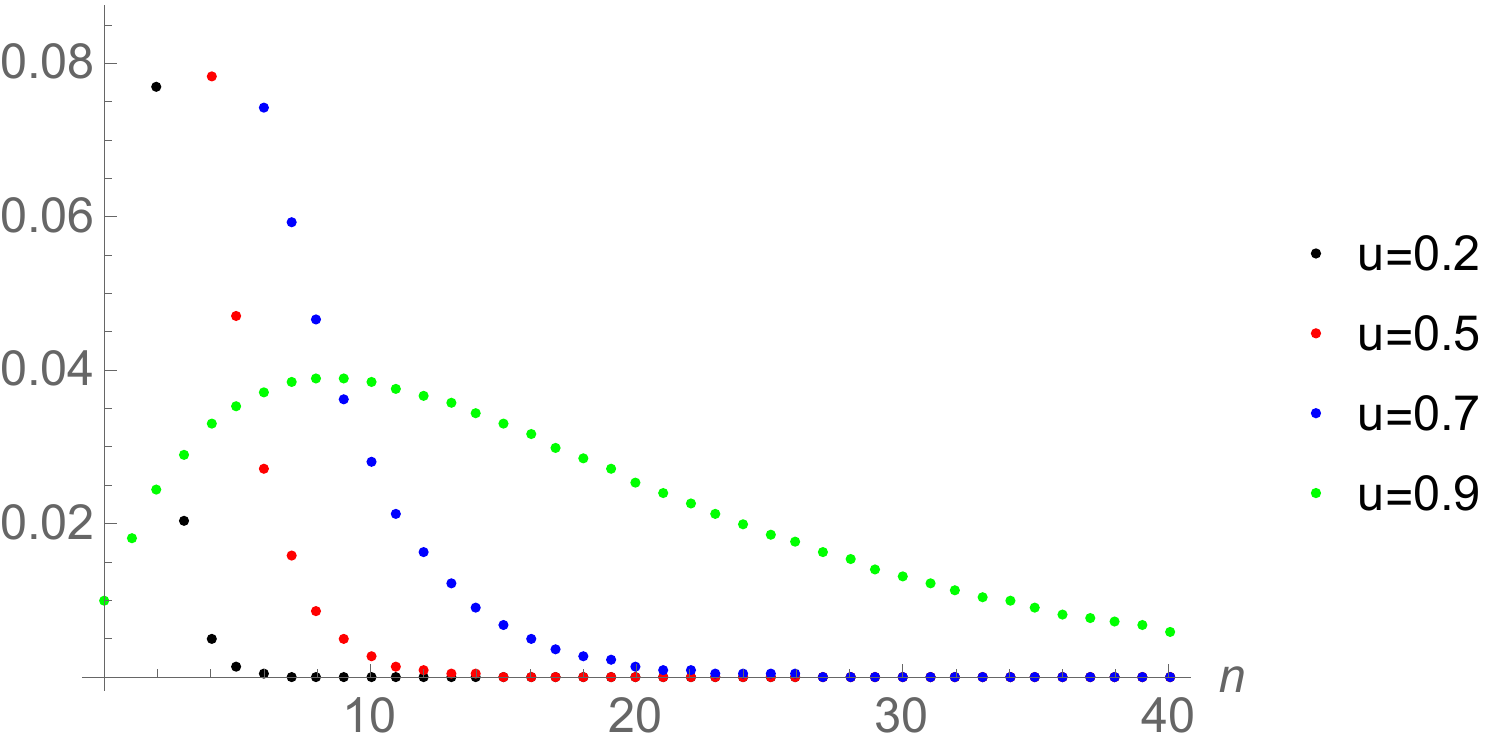}\quad
\includegraphics[width=0.45\textwidth]{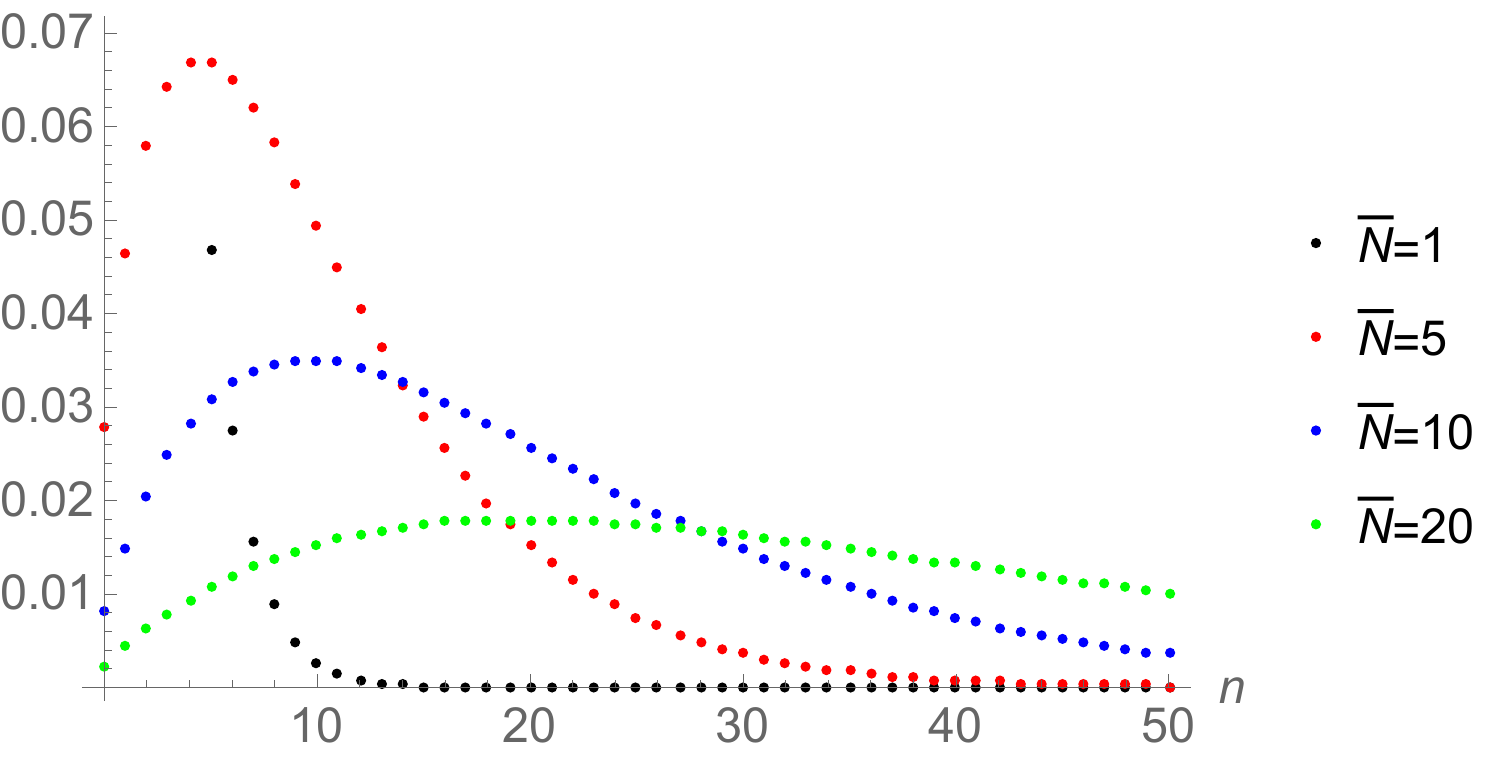}
{\footnotesize   }}
\caption{\footnotesize Graphs of $\sfP_n^{(1 ,0)}(u)$ (\ref{negbinomialdist})  and 
  $\widetilde{\sfP}^{(\varkappa ,0)}_n(\bar N)$ (\ref{negbinomialdist1}) for different values of $u$ and $\bar N$, respectively.} 
\label{figure1}
\end{figure}
%%%%%%%%%%%%%%%%%%%%%
%%%%%%%%%%%%%%%%%%%%%
In the limit $\eta= 1$, i.e., at the lowest bound $\vk = 1/2$ of the discrete series of SU$(1,1)$,  this distribution is reduced to the celebrated Bose-Einstein one for the thermal light . For $\eta< 1$, the difference might be understood from the fact that we consider the average photocount number $\bar N$ instead of the mean photon number $\bar n$ impinging on the detector in the same interval \cite{fox06}.  For a related interpretation within the framework of thermal equilibrium states of the oscillator see \cite{aharonov73}. 
Note that the above CS's, built from the  negative binomial distribution, were also discussed in \cite{algahel08}.  In  Fig.~\ref{figure1}.a we display the negative binomial  probability distribution  $n \mapsto \sfP_n^{(1 ,0)}(u)$ \eqref{negbinomialdist} for different values of $u$.  Note that as $u$ increases the maximum value of $\sfP_n^{(1 ,0)}(u)$ increases with $n$ and then  slowly decreases. A similar situation is displayed for $\widetilde{\sfP}^{(\varkappa ,0)}_n(\bar N)$ \eqref{negbinomialdist1} in Fig.~\ref{figure1}.b.

%%%%%%%%%%%%%%%%%%%%%
%%%%%%%%%%%%%%%%%%%%%
\begin{figure}[ht]
\centering
\subfigure[$\sfP^{(1,s)}_n(0.6)$, i.e. $\varkappa=1$ \hskip5cm (b) $\sfP^{(3,s)}_n(0.6)$,  i.e. $\varkappa=3$]
{\includegraphics[width=0.475\textwidth]{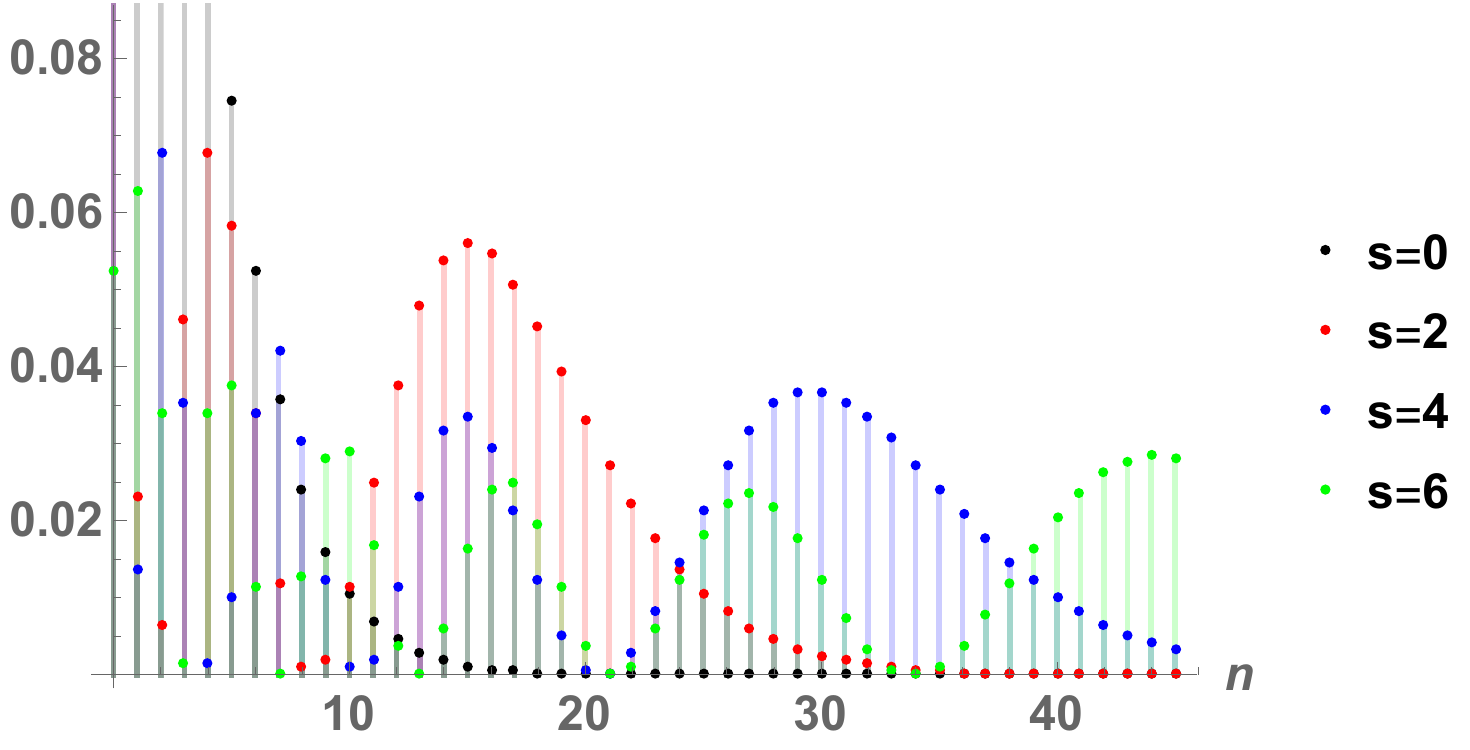}\qquad
\includegraphics[width=0.475\textwidth]{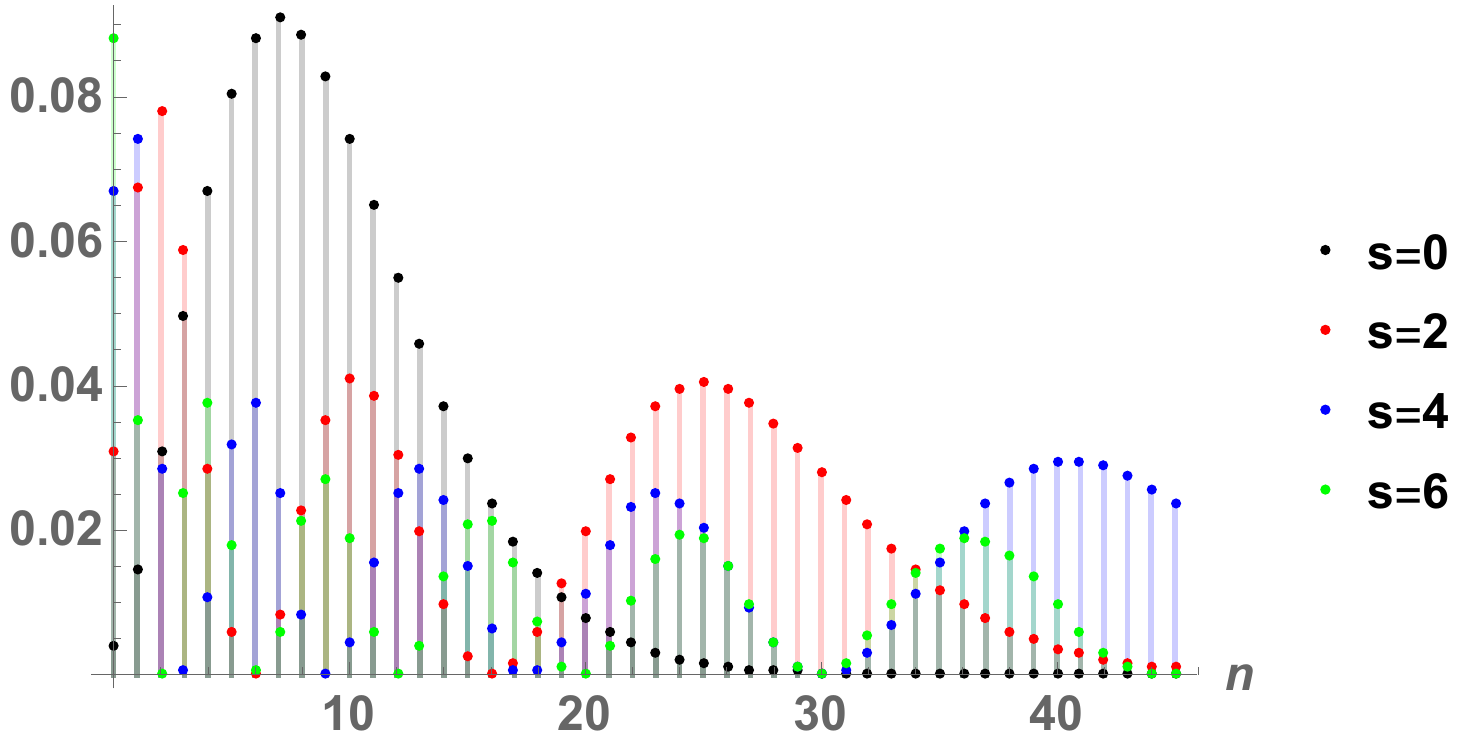}}
\caption{\footnotesize Graphs of $\sfP_n^{(\varkappa ,s)}(u)$ (\ref{detprobsss}) for different values of the pair 
$(\varkappa ,s)$ and $u=0.6$ in all the cases.}
\label{figure2}
\end{figure}
%%%%%%%%%%%%%%%%%%%%%
%%%%%%%%%%%%%%%%%%%%%

\subsection{Arbitrary \underline {s}}

In  the general case   $s\geq 0$ the CS  $|\alpha;\varkappa;s\rg$ are given by
\begin{equation}
\label{SU11CSs}
\begin{array}{lll}
&\hskip-0.75cm |\alpha;\varkappa;s\rg = 
 (1-\vert \alpha\vert^2)^{\varkappa}\\
&\hskip-0.75cm  \ds\times\left[
\sum_{n=0}^{s-1} \sqrt{\frac{n!\,\Gamma(2\varkappa + s)}{s!\,\Gamma(2\varkappa + n)}}\bar\alpha^{s-n}
\,P_n^{(s-n,2 \varkappa-1)} (1-2\vert\alpha\vert^2)\,|n\rg\right.\\[0.4cm]
&\hskip-0.75cm \left. \ds+\sum_{n\geq s} \sqrt{\frac{s!\,\Gamma(2\varkappa + n)}{n!\,\Gamma(2\varkappa + s)}}\alpha^{n-s}
\,P_s^{(n-s,2 \varkappa-1)} (1-2\vert\alpha\vert^2)\,|n\rg\right]\,.
\end{array}
\end{equation}
The corresponding probability detection $0\leq n\mapsto \sfP^{(\varkappa,s)}_n(u)$ is equal to
\begin{equation}\begin{array}{lll}
\label{detprobsss}
&\hskip-0.75cm  \sfP^{(\varkappa,s)}_n(u)=(1-u)^{2\varkappa}\times\\
&\hskip-0.75cm \times\left\{\begin{array}{ll}
 \ds \frac{n!\,\Gamma(2\varkappa + s)}{s!\,\Gamma(2\varkappa + n)}\,u^{s-n}
\,\left(P_n^{(s-n,2 \varkappa-1)} (1-2u)\right)^2\,,
& 
\;\;n<s\\[0.4cm]
\ds \frac{s!\,\Gamma(2\varkappa + n)}{n!\,\Gamma(2\varkappa + s)}\,u^{n-s}
\,\left( P_s^{(n-s,2 \varkappa-1)} (1-2 u)\right)^2 \,,&
\;\;n\geq s
\end{array}\right.\,.
\end{array}\end{equation}
In this case we get  for the expectation value of the number operator 
\begin{equation}
\label{barnQus}
\bar n^{\varkappa,s}\equiv \bar n= \frac{s+(s+2 \varkappa)\, u}{1-u} \quad \Leftrightarrow\quad
u=\frac{\bar n -s}{\bar n+s+2\varkappa}\,.
 \end{equation}
 In  Fig.~\ref{figure2} we display the  distribution  probability $n\mapsto \sfP_n^{(\varkappa ,s)}(u)$ \eqref{detprobsss} of detecting  photons 
  for different values of $\varkappa$ and $s\neq 0$ with a fixed value of $u=0.6$. Both graphs are for $s=0$ (no added photons).

 The Mandel parameter is given by
\begin{equation} \label{mandels}
Q^{(\varkappa,s)}(u) = \frac{2 (s^2+(2s+1)\varkappa)\,u}{(1-u)(s+(s+2\varkappa)\,u)}-1\,.
\end{equation}
Note  its limit value at $\varkappa\to \infty$:
\begin{equation} \label{mandels1}
\quad  Q^{(\infty,s)}(u) = \frac{2 s +u}{1-u}\,.
\end{equation}
Let us introduce the new parameter:
\begin{equation}
\label{paramr}
r:= \frac{s\eta}{1 + s\eta} \in [0,1)\, . 
\end{equation} 
This interesting parameter $r$, that combines $s$ and the previous  $\eta=1/(2\varkappa)$, is strictly increasing from $0$ to $1$ as $s\eta\in [0,\infty)$. However,  it cannot  be interpreted as an efficiency. The expressions of $Q$ and $\ud Q/\ud u$ are given in terms of $(r,s,u)$ by:
\begin{equation}
\label{Qsur}
Q= \frac{u^2 + 2su -r}{(1-u)(u+r)}\, , \quad \frac{\ud Q}{\ud u}= \frac{(2s +1-r)u^2 +2sr +r(1-r)}{(1-u)^2(u+r)^2}\, . 
\end{equation}
Hence, $Q$ is strictly increasing from  $0$ if $s=0$ and from $-1$ if $s>0$,  to $\infty$ at $u=1$. For $s>0$, coherent states 
are sub-Poissonian for $0\leq u< \sqrt{s^2 +r} - s$, are Poissonian at $u= \sqrt{s^2 +r} - s$, and become 
super-Poissonian at $ \sqrt{s^2 +r} - s< u <1$.
%%%%%%%%%%%%%%%%%%%%%
%%%%%%%%%%%%%%%%%%%%%
\begin{figure}[ht]
\centering
\subfigure[$Q^{(1,s)}(u)$ \hskip5.5cm (b) Detail of the negative part of $Q^{(1,s)}(u)$]{
\includegraphics[width=0.45\textwidth]{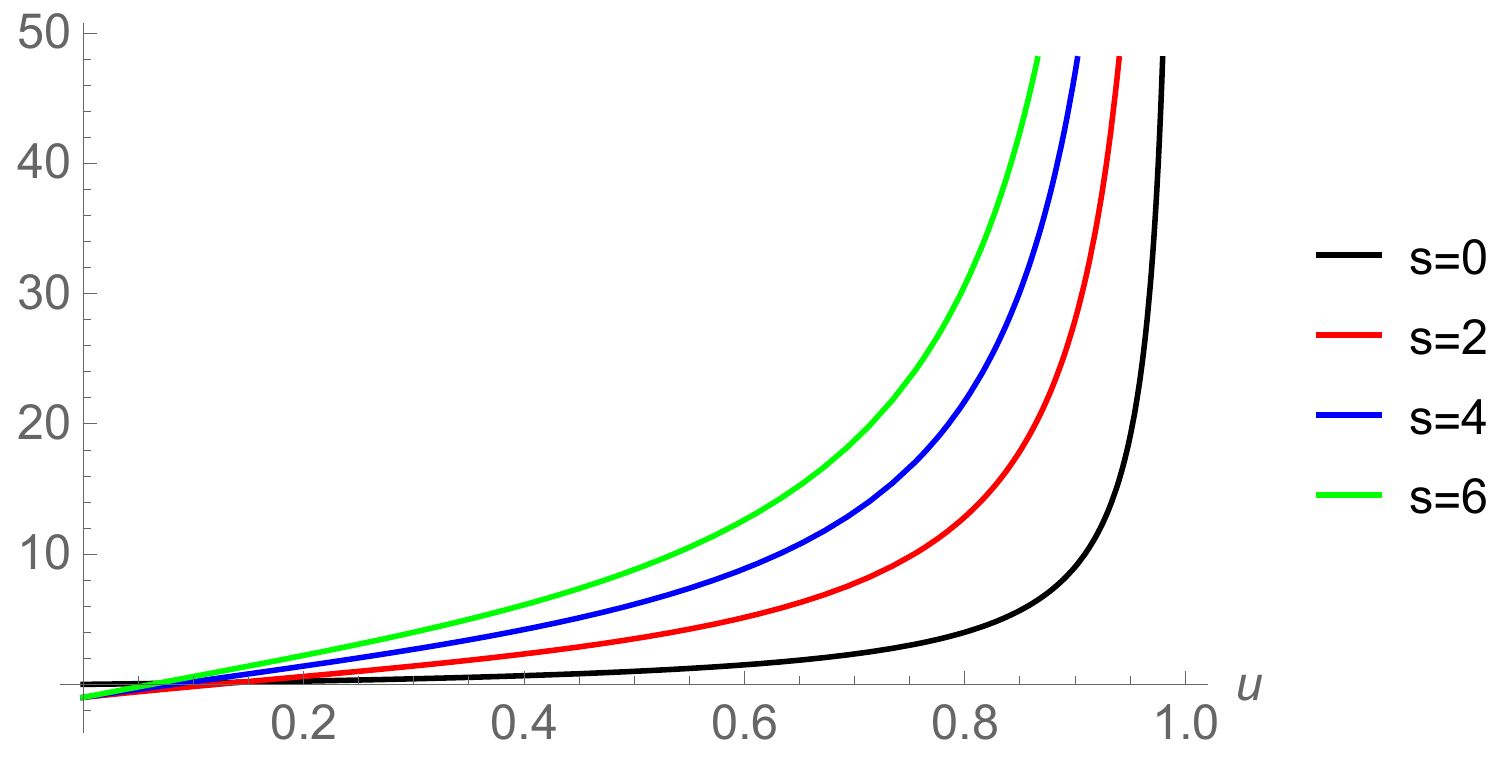}\qquad  
\includegraphics[width=0.45\textwidth]{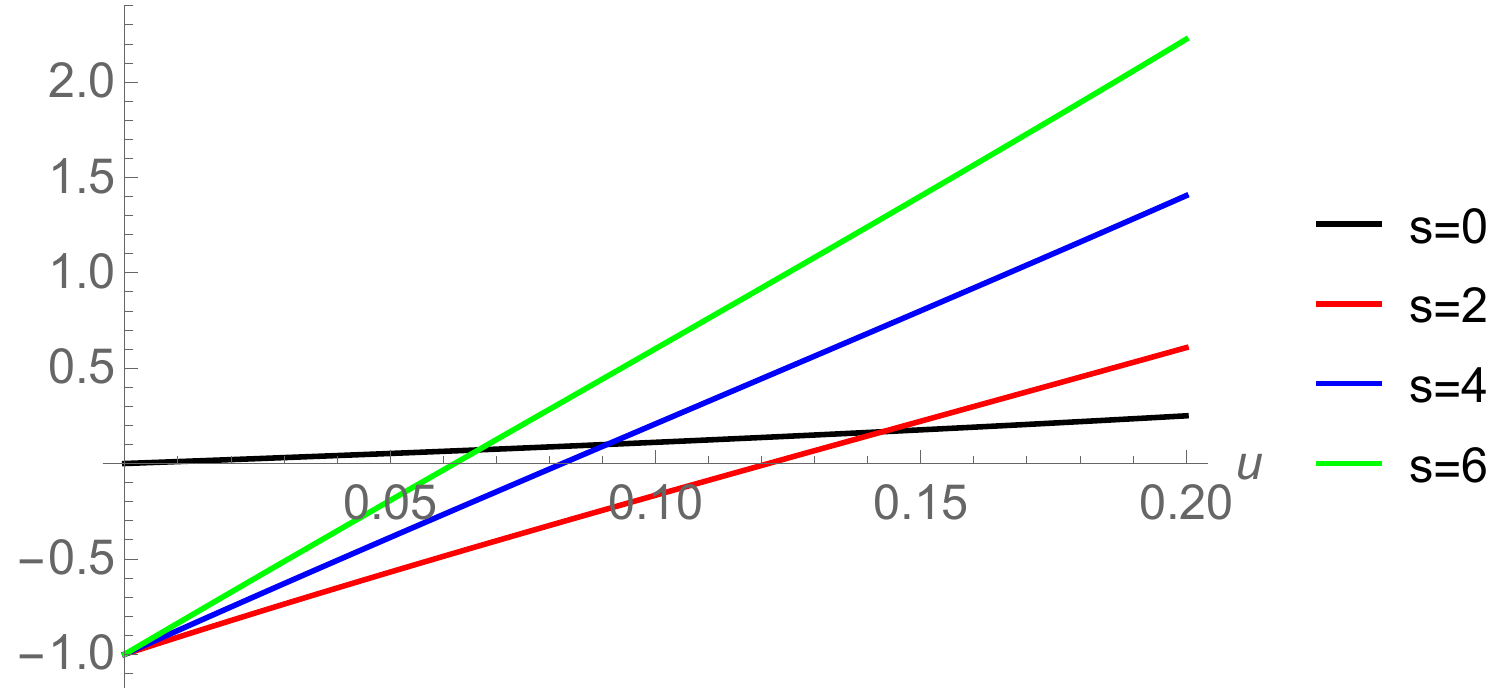}
}
\caption{\footnotesize Graphs of the Mandel parameter $Q^{(\varkappa,s)}(u)$ (\ref{mandels})   for $\varkappa=1$ and different values of $s$.}
\label{figure3}
\end{figure}\medskip

%%%%%%%%%%%%%%%%%%%%%
%%%%%%%%%%%%%%%%%%%%%
\begin{figure}[ht]
\centering
\subfigure[$Q^{(\varkappa,1)}(u)$\hskip5.5cm Detail of the negative part of $Q^{(\varkappa,1)}(u)$]{
\includegraphics[width=0.45\textwidth]{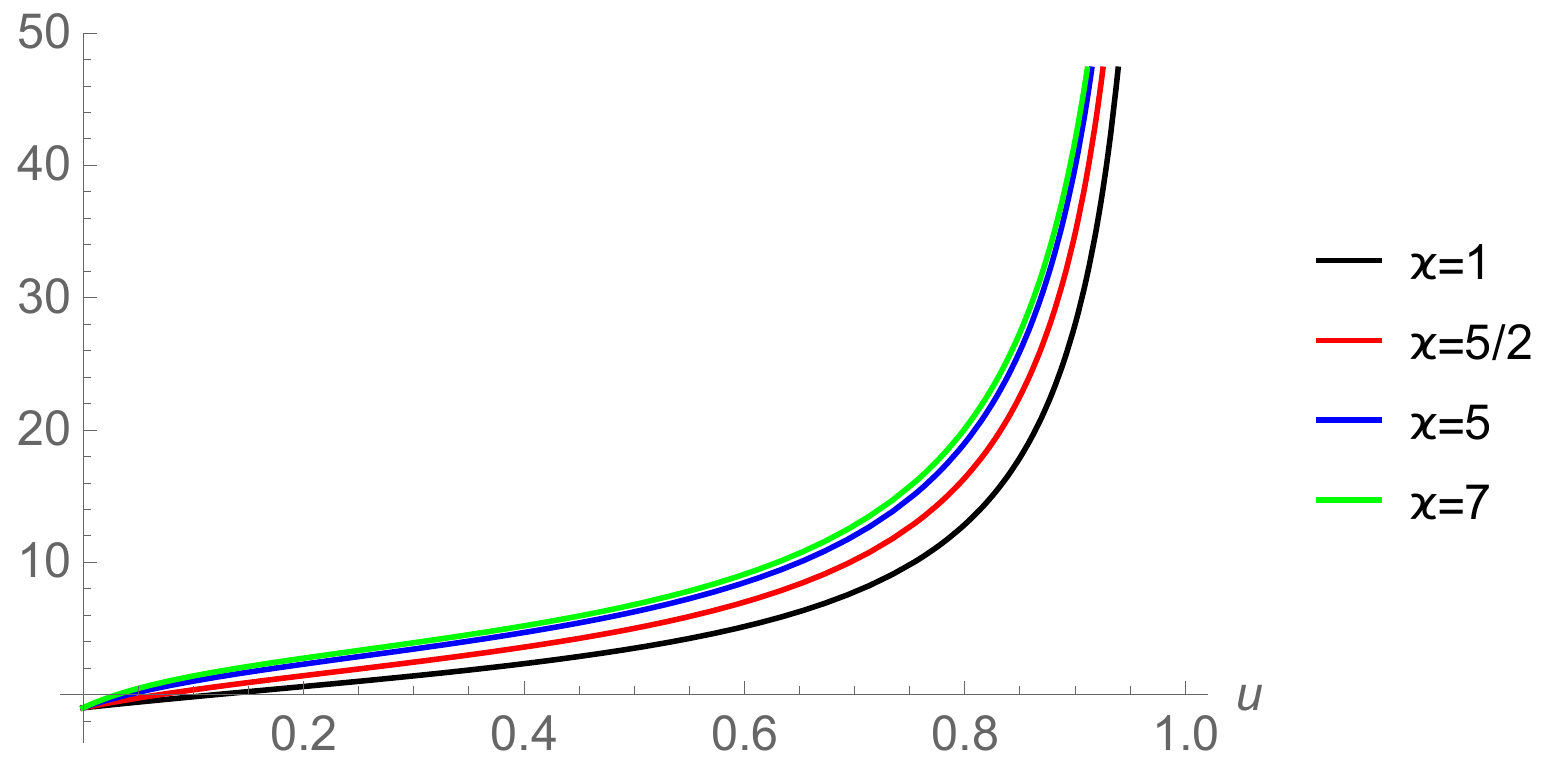}\qquad  
\includegraphics[width=0.45\textwidth]{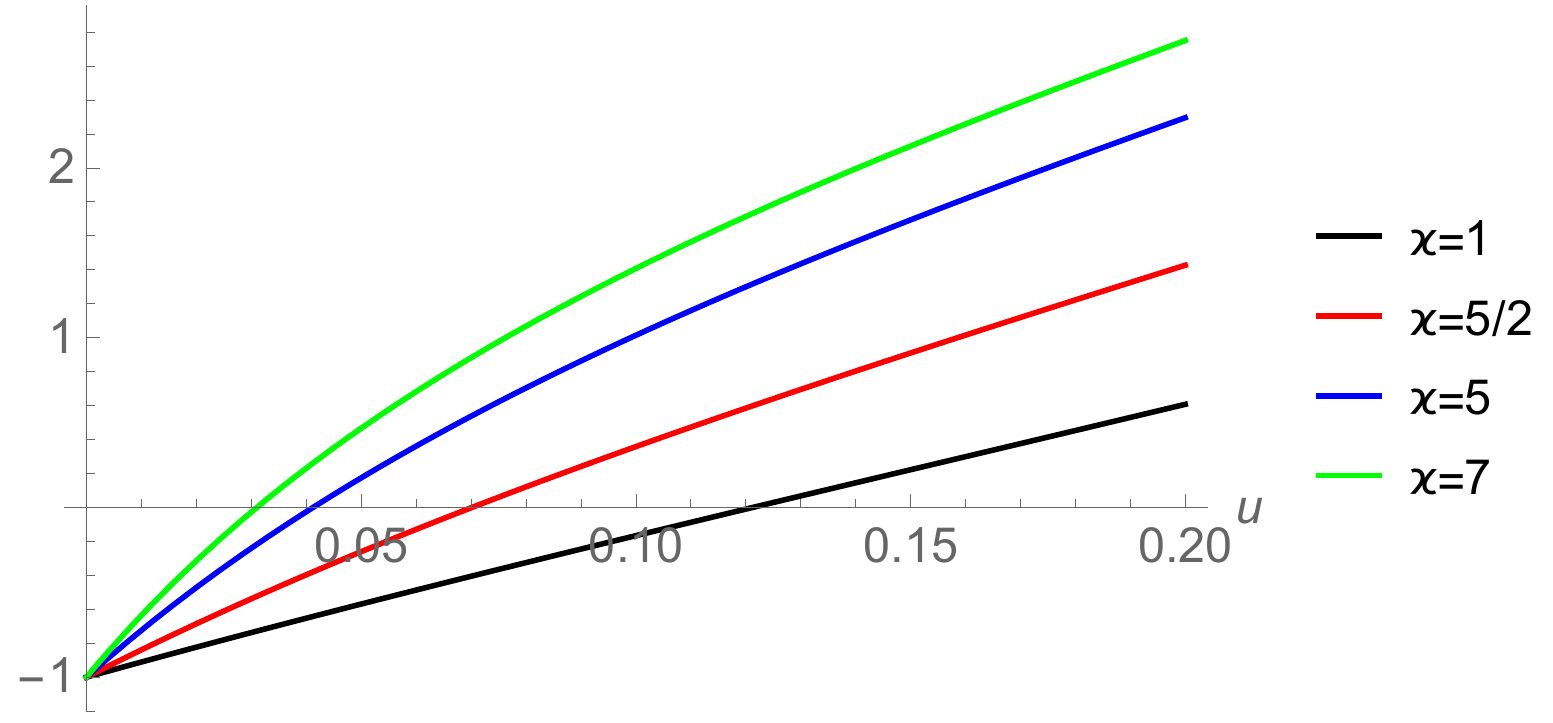}}
\caption{\footnotesize Graphs of the Mandel parameter $Q^{(\varkappa,s)}(u)$ (\ref{mandels})   for different values of  $\varkappa$ and for $s=1$.}
\label{figure4}
\end{figure}
%%%%%%%%%%%%%%%%%%%%%
%%%%%%%%%%%%%%%%%%%%%
\begin{figure}[h]
\centering
\includegraphics[width=0.475\textwidth]{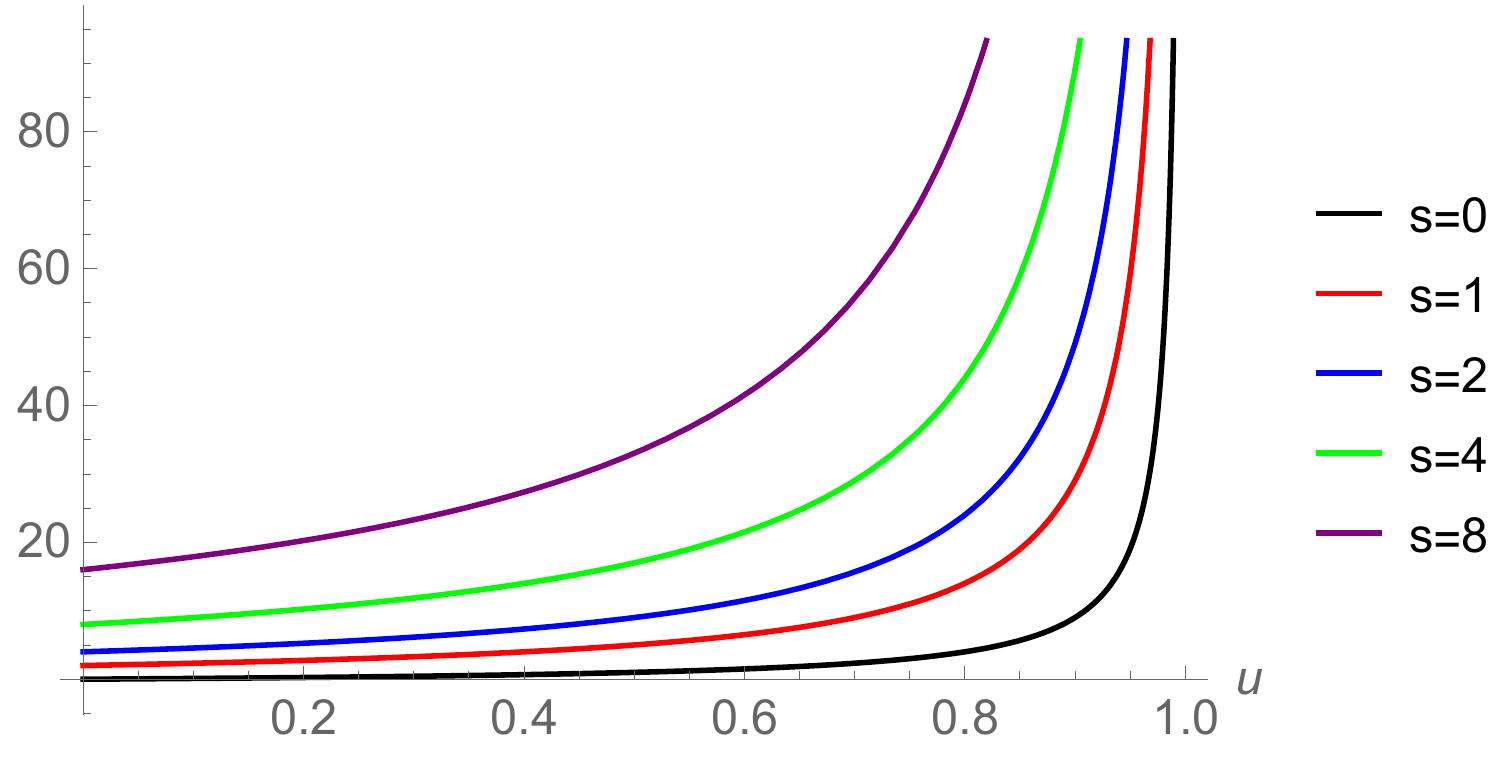}
\caption{\footnotesize  Graphs of the Mandel parameter $Q^{(\infty,s)}(u)$ (\ref{mandels1})   for  different values of $s$  and $\varkappa=\infty$.}\label{figure5}
\end{figure}
%%%%%%%%%%%%%%%%%%%
%%%%%%%%%%%%%%%%%%%

In Figs.~\ref{figure3}, \ref{figure4}, \ref{figure5} we represent the Mandel parameter \eqref{mandels} for different values of $\varkappa$ and $s$. In Fig.~\ref{figure3}.a we display $Q^{(\varkappa,s)}(u)$ with $\varkappa=1$ for different values of $s$, and in Fig.~\ref{figure4}.a we display $Q^{(\varkappa,s)}(u)$ with $s=1$, and  for different values of $\varkappa$. In Fig.~\ref{figure3}.b and Fig.~\ref{figure4}.b we consider the subinterval of $u$ corresponding to the  sub-Poissonian regime, i.e. $Q<0$. Finally in Fig.~\ref{figure5} we consider the limit $\varkappa = \infty$ where the regime is super-Poissonian, i.e. $Q>0$ for any value of $s$.

Finally, the unitary operator $U^{\varkappa}(p( \bar \alpha))$, introduced  in \eqref{morecssu11} to build the SU$(1,1)$-CS's, can  be  expressed in terms  the generators $K_{\pm}$ for the representation $U^{\varkappa}$ \cite{perel86} as  
\begin{equation}
\label{SU11disp}
U^{\varkappa} (p(\bar \alpha)) = e^{\varrho_{\alpha}\, K_+ - \bar\varrho_{\alpha} \, K_-}\, ,
\end{equation}
where 
\begin{equation}
\label{disp}
\varrho_{\alpha}= \tanh^{-1}\vert \alpha\vert\,e^{\ii \arg \alpha}\,,\quad
\alpha =  \tanh(\vert \varrho_{\alpha}\vert) \,e^{\ii \arg \alpha}\,,
\end{equation} 
with the actions
\begin{align}
\label{raisstaFBsu11}
K_+ |n\rg &= \sqrt{(n+1)(2 \varkappa + n)}\, |n+1\rg\, ,\\ 
\label{lowstaFBsu11}K_- |n\rg  &= \sqrt{n \, (2 \varkappa + n-1)}\, |n-1\rg\\ 
\label{K0FBsu11}
[K_-,K_+]&=2K_0\, , \quad K_0|n\rg = (n+\varkappa)|n\rg\, .
 \end{align}
The expression  \eqref{SU11disp} is reminiscent of the displacement operator $D(\alpha)=e^{\alpha a^{\dag}-\overline{\alpha}a}$ whose action on the vacuum yields the standard CS's. In fact, it can be viewed as a unitary ``displacement'' for the Lobatchevsky or hyperbolic geometry of the unit disk. 

Thus, we have the meaningful formulae relating the complex parameter  $\varrho_{\alpha}$ 
\eqref{disp} to the optical phase space parameter \eqref{mapaln}:
\begin{equation}
\label{rhoxial}
 \xi_{\alpha}= \sqrt{\bar n}\,e^{\ii \arg \alpha}= \sqrt{(\varkappa +s)\,\cosh2\vert\varrho_{\alpha}\vert-\varkappa}\, e^{\ii \arg \alpha}\,. 
\end{equation}
\subsection{Squeezing}
As asserted in Ref.\;\cite{wodeb85} two-photon processes involved in nonlinear parametric generations or two-photon conversion can be modelled with the effective Hamiltonian (in a rotating wave approximation):
\begin{equation}
\label{2phot1}
H=\ii \frac{\lambda\mathcal{E}}{2}\left(a^2-{\adg}^2\right)\, , 
\end{equation}
where $\mathcal{E}$ is the external electric-pump generator and $\lambda$ is the proper coupling constant of the pump to the nonlinear medium described by the operators \eqref{acaadag}. Let us now choose for $\varkappa$  the value $\varkappa=1/4$ (resp. $\varkappa=3/4$). Note that  the overcompleteness \eqref{runitmorecssu11} does not make sense for the former value. Let us re-label  the  number states   in the original  Fock-Hilbert space $\mathcal{H}$ as $|n\rg= \left| \frac{m}{2}\right\rg\equiv |m\rg_{\mathrm{e}}$, for even $m\in \N$  (resp. $|n\rg= \left| \frac{m-1}{2}\right\rg\equiv |m\rg_{\mathrm{o}}$, for odd $m\in \N$).  When viewing the operators $K_{\pm}$, $K_0$, $a$, $\adg$ in \eqref{raisstaFBsu11}, \eqref{lowstaFBsu11} and \eqref{K0FBsu11}  as acting in $\mathcal{H}_{\mathrm{e}}$ (resp. $\mathcal{H}_{\mathrm{o}}$) spanned by the $|m\rg_{\mathrm{e}}$'s (resp. $|m\rg_{\mathrm{o}}$'s), one obtains:
\begin{equation}
\label{}
K_- = \frac{1}{2}a^2\, , \ K_+ = \frac{1}{2}{\adg}^2\,, \ K_0= \frac{1}{2}\left(a\adg + \frac{1}{2}\right)\,. 
\end{equation}
Hence, for these two values of $\varkappa$, and with this reinterpretation of the number states in the original Fock space, the unitary action \eqref{SU11disp} of $U^{\varkappa} (p(\bar \alpha))$ on number  states $[s\rg$ is the same as for the construction of squeezed states \cite{stoler70}, and we can put \eqref{2phot1} under the form: 
\begin{equation}
\label{2phot2 }
H=\ii \lambda\mathcal{E}\left(K_- - K_+\right)\, . 
\end{equation}
  Let us now explore in detail the squeezing properties of the states  $|\alpha;\varkappa;s\rg$ given in \eqref{SU11CSs}. By using some material from Ref. \cite{gazolmo18-1}   the expectation values and variances, in these states,  of the generators $K_0$, $K_1 = \frac{\ii}{2}(K_+ - K_-)$ and $K_2= \frac{1}{2}(K_+ + K_-)$, with $K_0$, $K_{\pm}$ given in \eqref{K0FBsu11}, \eqref{raisstaFBsu11}, and \eqref{lowstaFBsu11} respectively, are given by:
 \begin{equation}\label{mean0pm} \begin{array}{lll}
  \lg K_0\rg   &=&\ds (\varkappa +s)\frac{1+\vert\alpha\vert^2}{1-\vert\alpha\vert^2}\, , \\[0.3cm]
      \lg K_{\pm}\rg    &=& \ds\frac{(\varkappa +s)}{1-\vert\alpha\vert^2}\, \alpha_{\pm}\,,
      \qquad \alpha_+=\alpha,\,
      \alpha_-=\bar \alpha,\, \\[0.4cm]
\lg K_1\rg &=&\ds- (\varkappa +s)\frac{\mathrm{Im}\,\alpha}{1-\vert\alpha\vert^2}\, ,   \\[0.3cm]
\lg K_2\rg &=&\ds  (\varkappa +s)\frac{\mathrm{Re}\,\alpha}{1-\vert\alpha\vert^2}\, ,\\[0.4cm]
 \Delta K_1  &=&\ds\sqrt{\frac{\varkappa + s(s+2\varkappa) }{2}} \frac{\vert1- \alpha^2\vert}{1-\vert\alpha\vert^2}\, , \\[0.3cm]
  \Delta K_2  &=&\ds\sqrt{\frac{\varkappa + s(s+2\varkappa)}{2}} \frac{\vert1+ \alpha^2\vert}{1-\vert\alpha\vert^2} \,. 
\end{array}\end{equation}
From the commutation relation $[K_1,K_2]=-\ii K_0$ one derives the inequality 
\begin{equation}
\label{ineq}
 \Delta K_1  \, \Delta K_2 \geq  \frac{1}{2}\vert  \lg K_0\rg\vert\,,
\end{equation}
and, from \eqref{mean0pm},  its explicit form in terms of $\alpha$, $s$, and $\varkappa$ becomes   
\begin{equation}
\label{ineqsk}
(\varkappa + s(s+2\varkappa)) \vert 1 -\alpha^4\vert  \geq  (\varkappa +s) (1-\vert\alpha\vert^4)\,.
\end{equation}
With  the parametrisation $\alpha = \tanh \vert\varrho_{\alpha}\vert \,e^{\ii\phi}$ introduced in \eqref{SU11disp}, one obtains from \eqref{ineqsk} the trivial inequality:
\begin{equation}
\label{inethtephi}
\sqrt{1 + \frac{1}{4}(\tanh 2\vert\varrho_{\alpha}\vert)^2\, (\sinh 2 \vert\varrho_{\alpha}\vert)^2\sin^42\phi}\geq \frac{\varkappa +s}{\varkappa + s(s+2\varkappa)}\,. 
\end{equation}
Note the relation $\varrho_{\alpha} = -\lambda\mathcal{E}$ between this parametrisation and the physical one in the cases $\varkappa=1/4$ or $\varkappa=3/4$.

%%%%%%%%%%%%%%%%%%%%%
%%%%%%%%%%%%%%%%%%%%%
\begin{figure}[h]
\centering
\includegraphics[width=0.475\textwidth]{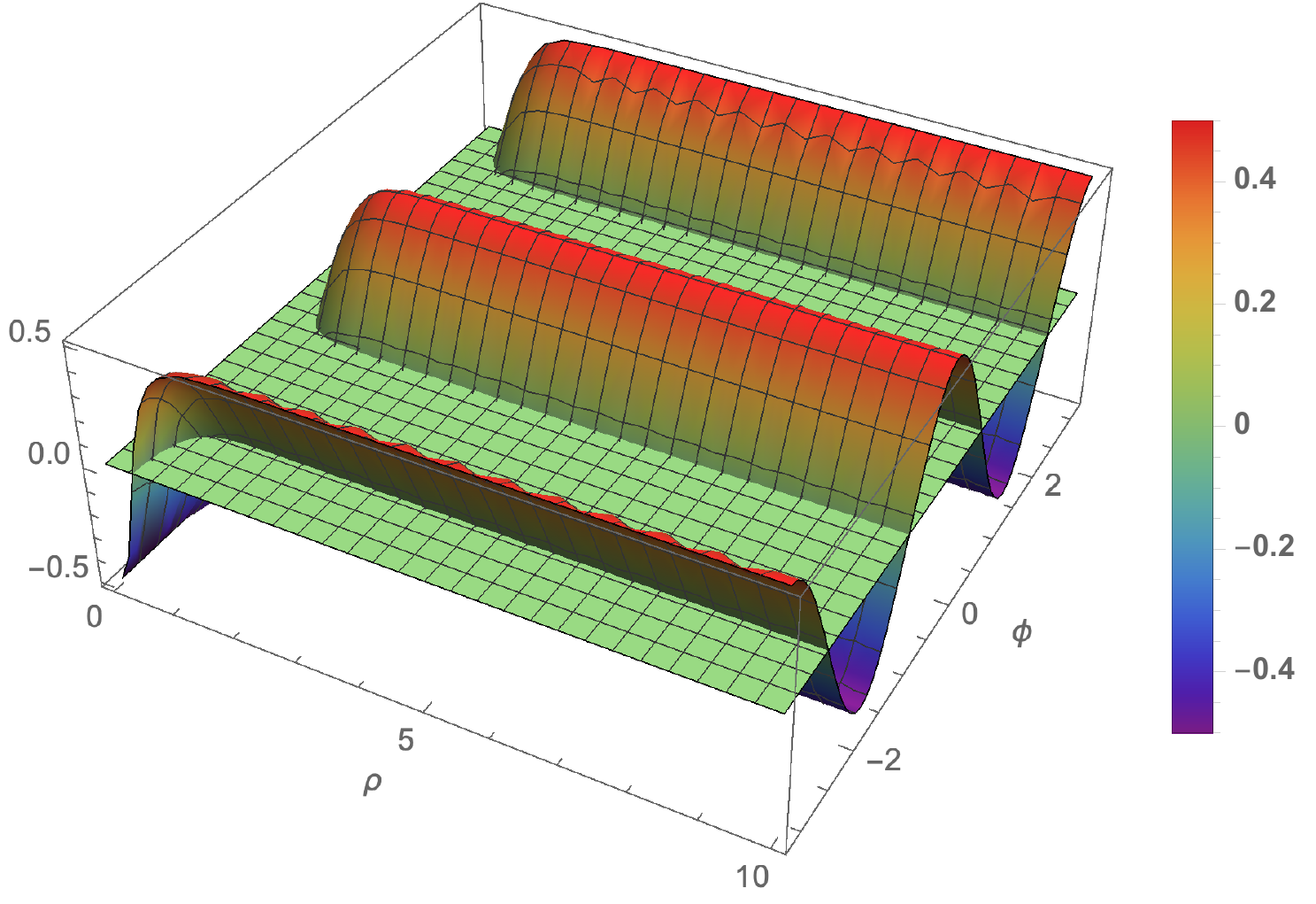}
\caption{\footnotesize Graph of the inequality  (\ref{sqeez1+}) for $\varkappa=1$ and $s=0$. Actually we represent $\Delta=S -(1-  \tanh^2 2\vert\varrho_{\alpha}\vert\,\cos^2\phi)$. The colorbar shows this difference.}\label{figure6}
\end{figure}
%%%%%%%%%%%%%%%%%%%
%%%%%%%%%%%%%%%%%%%

%%%%%%%%%%%%%%%%%%%%%
%%%%%%%%%%%%%%%%%%%%%
\begin{figure}[h]
\centering
\includegraphics[width=0.475\textwidth]{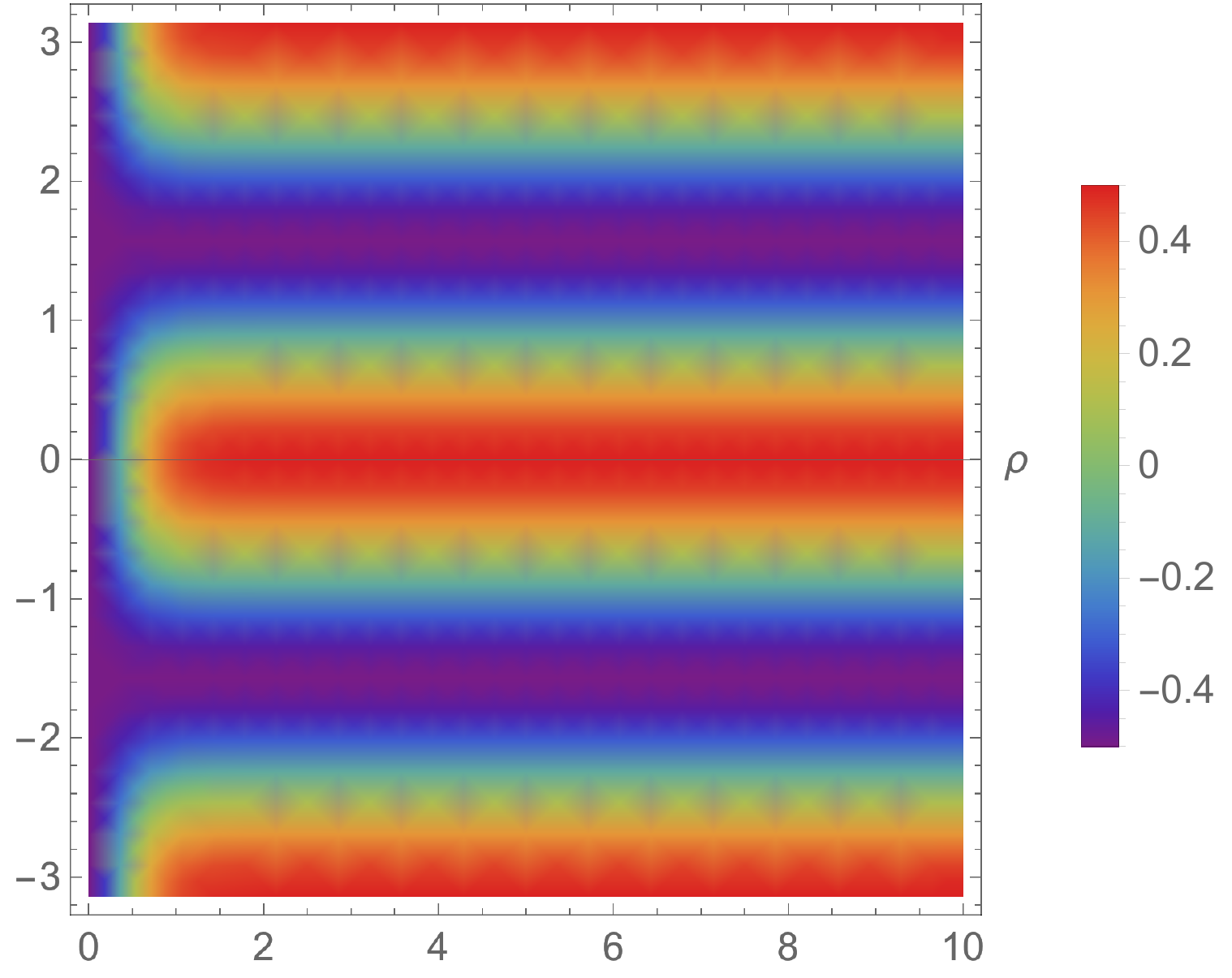}
\caption{\footnotesize Density plot  of the difference $\Delta=S -(1-  \tanh^2 2\vert\varrho_{\alpha}\vert\,\cos^2\phi)$ (\ref{sqeez1+}) for $\varkappa=1$ and $s=1$. The colorbar shows the variation of $\Delta$.}
\label{figure7}
\end{figure}
%%%%%%%%%%%%%%%%%%%
%%%%%%%%%%%%%%%%%%%

We  now examine the  inequalities describing possible squeezing for the ``SU(1,1)-quadratures'' $K_1$ or $K_2$.  In our sense, and contrarily to the definition given in \cite{wodeb85}, squeezing holds when  either $\Delta K_1\leq \frac{1}{2}\vert  \lg K_0\rg\vert$ or $\Delta K_2\leq \frac{1}{2}\vert  \lg K_0\rg\vert$, while \eqref{ineq} remains true, of course. Hence, we obtain from the squares of  these exclusive alternatives
\begin{align}
\label{sqeez1+} \mathrm{either}\quad S & \geq 1-  \tanh^2 2\vert\varrho_{\alpha}\vert\,\cos^2\phi\, , \\
\label{sqeez1-}  \mathrm{or} \quad   S & \geq 1- \tanh^2 2\vert\varrho_{\alpha}\vert\,\sin^2\phi\,, 
\end{align}
with $S:= \dfrac{1}{2}\dfrac{(\varkappa +s)^2}{\varkappa +s(s+2\varkappa)}$.

Figs.~\ref{figure6} and \ref{figure7}  display  the inequality  \eqref{sqeez1+} in the form 
$\Delta=S -( 1-  \tanh^2 2\vert\varrho_{\alpha}\vert\,\cos^2\phi )$ for different values of $\varkappa$ and $s$. The  3D plot in Fig.~\ref{figure6} corresponds to  the case  $s=0$ with $\Delta$ on the vertical axis.  In Fig.~\ref{figure7}  we alternatively  present a density plot where the colorbar  corresponds to the variation of  $\Delta$.

With the definition given in \cite{wodeb85} we would instead get:
\begin{align}
\label{sqeez2+} \mathrm{either}\quad S^{\prime}\cosh 2\vert\varrho_{\alpha}\vert & \geq 1 + \sinh^2 2\vert\varrho_{\alpha}\vert\,\sin^2 \phi\, , \\
\label{sqeez2-}  \mathrm{or} \quad   S^{\prime}\cosh 2\vert\varrho_{\alpha}\vert & \geq 1 + \sinh^2 2\vert\varrho_{\alpha}\vert\,\cos^2 \phi\, , 
\end{align}
with $S^{\prime}:= \dfrac{\varkappa +s}{\varkappa +s(s+2\varkappa)}$.

%%%%%%%%%%%%%%%%%%%%%%%%%%%%%%%%%%%%%%%%%%%%%%%%%%%%%%
%%%%%%%%%%%%%%%%%%%%%%%%%%%%%%%%%%%%%%%%%%%%%%%%%%%%%%
\section{SU$(1,1)$-CS quantization} \label{ANCSQsection}

\subsection{The quantization map and its complementary}
Let us consider  the resolution of the identity \eqref{resid} provided by a  family of non-standard CS determined by the sequence of functions $\boldsymbol{\mathsf{h}}:=\left(h_n(u)\right)$ \eqref{hnu}. This property makes the quantisation of functions (or distributions) $f(\alpha)$ possible along the linear map
\begin{equation}
\label{ANCSquant}
f(\alpha) \mapsto A^{\bsh}_f= \int_{\vert \alpha\vert < R}\frac{\ud^2\alpha}{\pi} \,w^{\bsh}(\vert \alpha\vert^2)\, f(\alpha)\,|\alpha\rg\lg\alpha|\,,  
\end{equation}
where we have written $|\alpha\rg= |\alpha; \varkappa;s\rg$ for short. 
The expectation value of $A^{\bsh}_f$ in the same CS's  provides the  ``semi-classical'' optical phase space portrait, or \textit{lower symbol},  of this operator:
\begin{equation*}
\label{phspport}
\lg\alpha|A^{\bsh}_f|\alpha\rg = \int_{\vert \beta\vert < R}\frac{\ud^2\beta}{\pi} \,w^{\bsh}(\vert \beta\vert^2)\, f(\beta)\,\vert\lg\alpha|\beta\rg\vert^2\equiv \widecheck{f^{\bsh}}(\alpha)\,. 
\end{equation*}
Let us fix $\alpha$. Due to \eqref{ANCSquant} the map  
\begin{equation}\label{portrait}
\beta \mapsto w^{\bsh}(\vert \beta\vert^2)\, \vert\lg\alpha|\beta\rg\vert^2= \sfP^{\bsh}_{\alpha}(\beta)
\end{equation}
 is a probability distribution on the centered disk $\mathcal{D}_{R}$ of radius $R$. Hence, the map $f(\alpha) \mapsto \widecheck{f^{\bsh}}(\alpha)$ is a local, generally regularising, averaging, of the original $f$.  

%%%%%%%%%%%%%%%%%%
%%%%%%%%%%%%%%%%%%
\begin{figure}[ht]
\centering
\subfigure[$\sfP^{\bsh}_{\alpha}(\beta)$ with   $s=0$\hskip3cm (b) $\sfP^{\bsh}_{\alpha}(\beta)$ with   $s=1$ ]{
\includegraphics[width=0.45\textwidth]{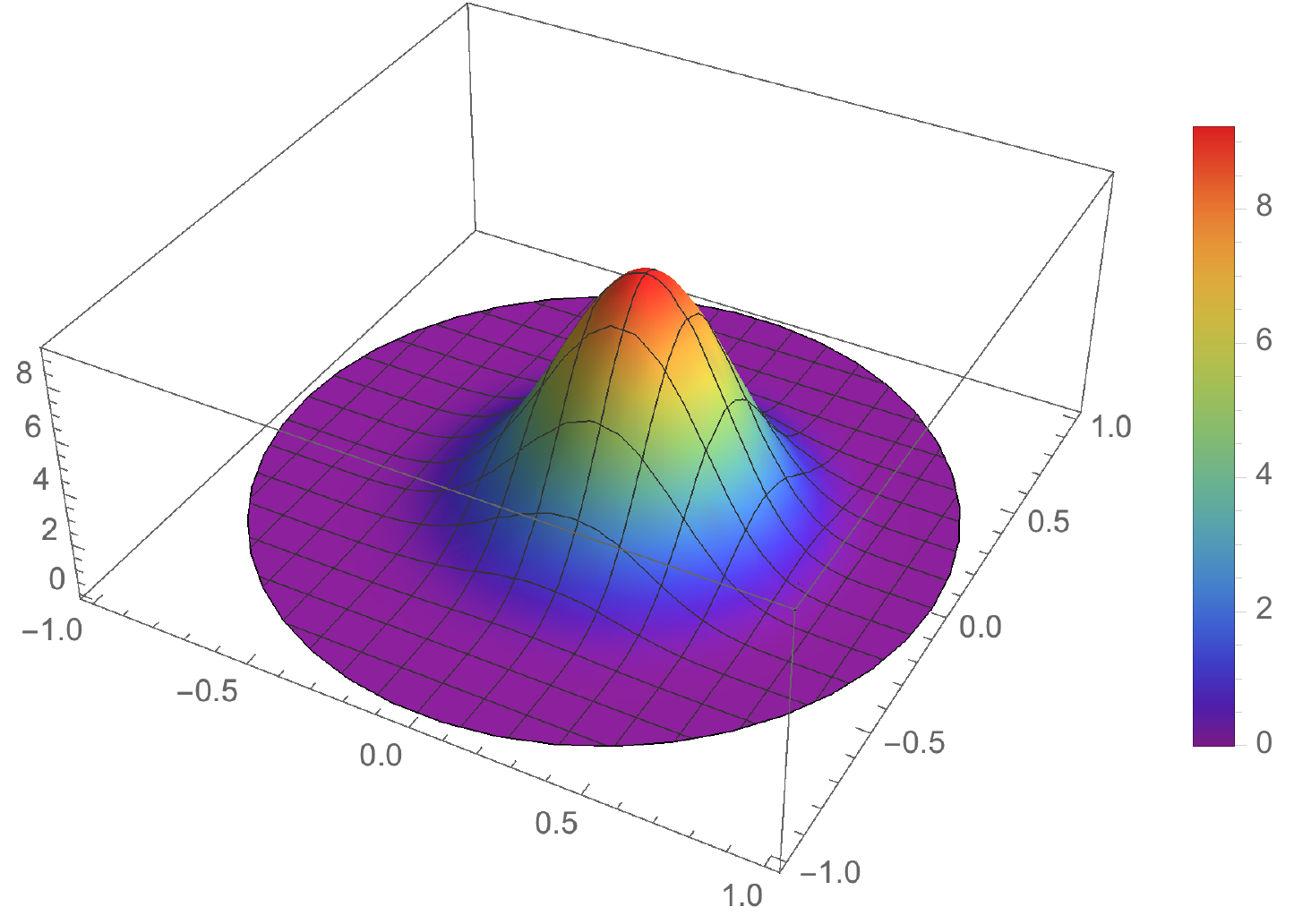} \quad
\includegraphics[width=0.45\textwidth]{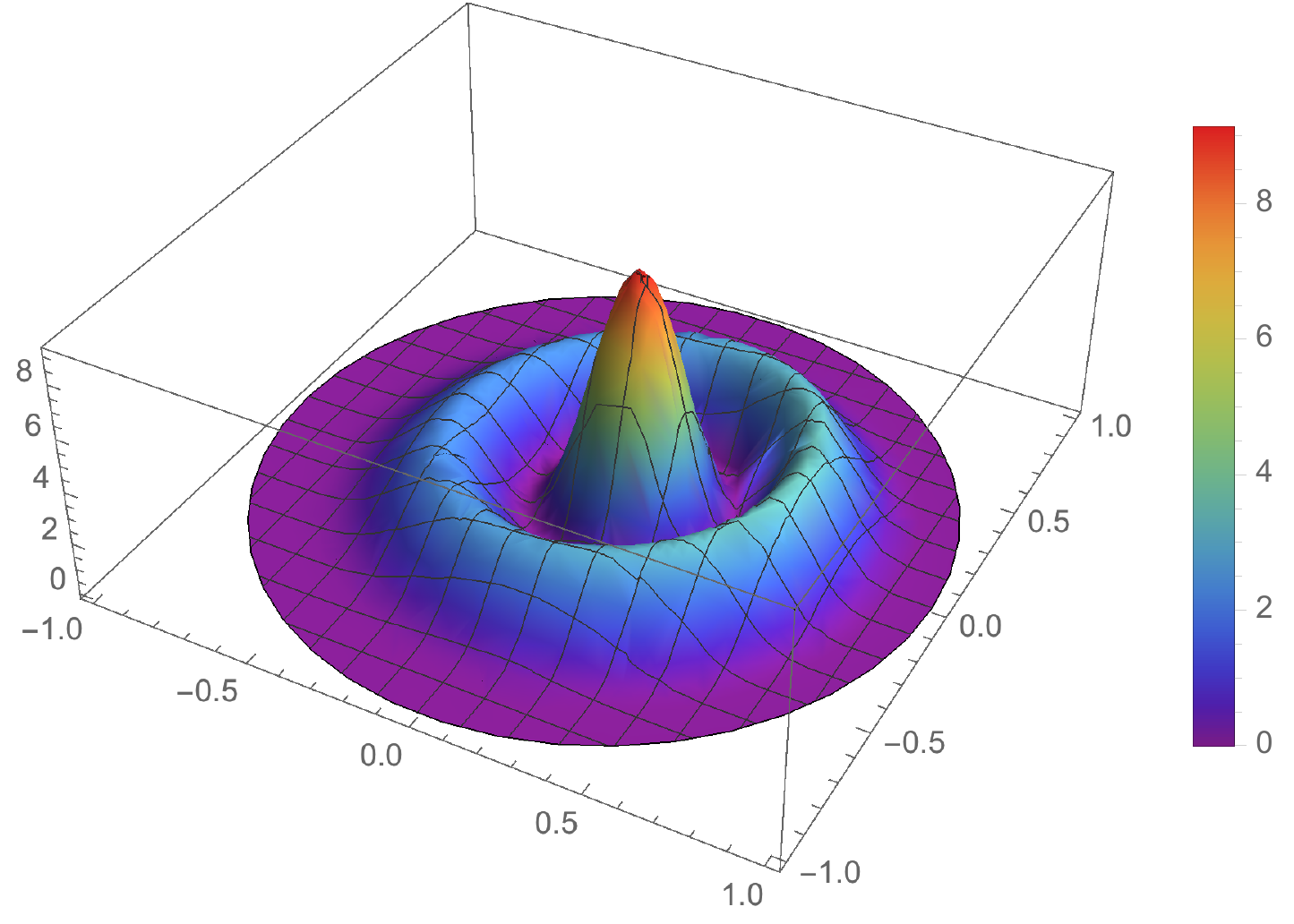}}
\caption{\footnotesize Graphs of $\sfP^{\bsh}_{\alpha}(\beta)$ (\ref{portrait}) on the unit disk background  for  $\alpha=0.1$,   
  $\varkappa=5$ and different values of $s$.  The colorbar shows the values on the vertical axis. } 
\label{figure8}
\end{figure}
%%%%%%%%%%%%%%%%%%%%%%%%%%%%%%%%%%%%
%%%%%%%%%%%%%%%%%%%%%%%%%%%%%%%%%%%%

The quantization map \eqref{ANCSquant} can be extended to cases  comprising geometric constraints in the optical phase portrait through the map \eqref{mapaln}, and encoded by distributions like Dirac or Heaviside functions. 

\begin{figure}[ht]
\centering
\subfigure[$\sfP^{\bsh}_{\alpha}(\beta)$ with $s=2$\hskip4cm (b) $\sfP^{\bsh}_{\alpha}(\beta)$ with $s=5$]{
\includegraphics[width=0.475\textwidth]{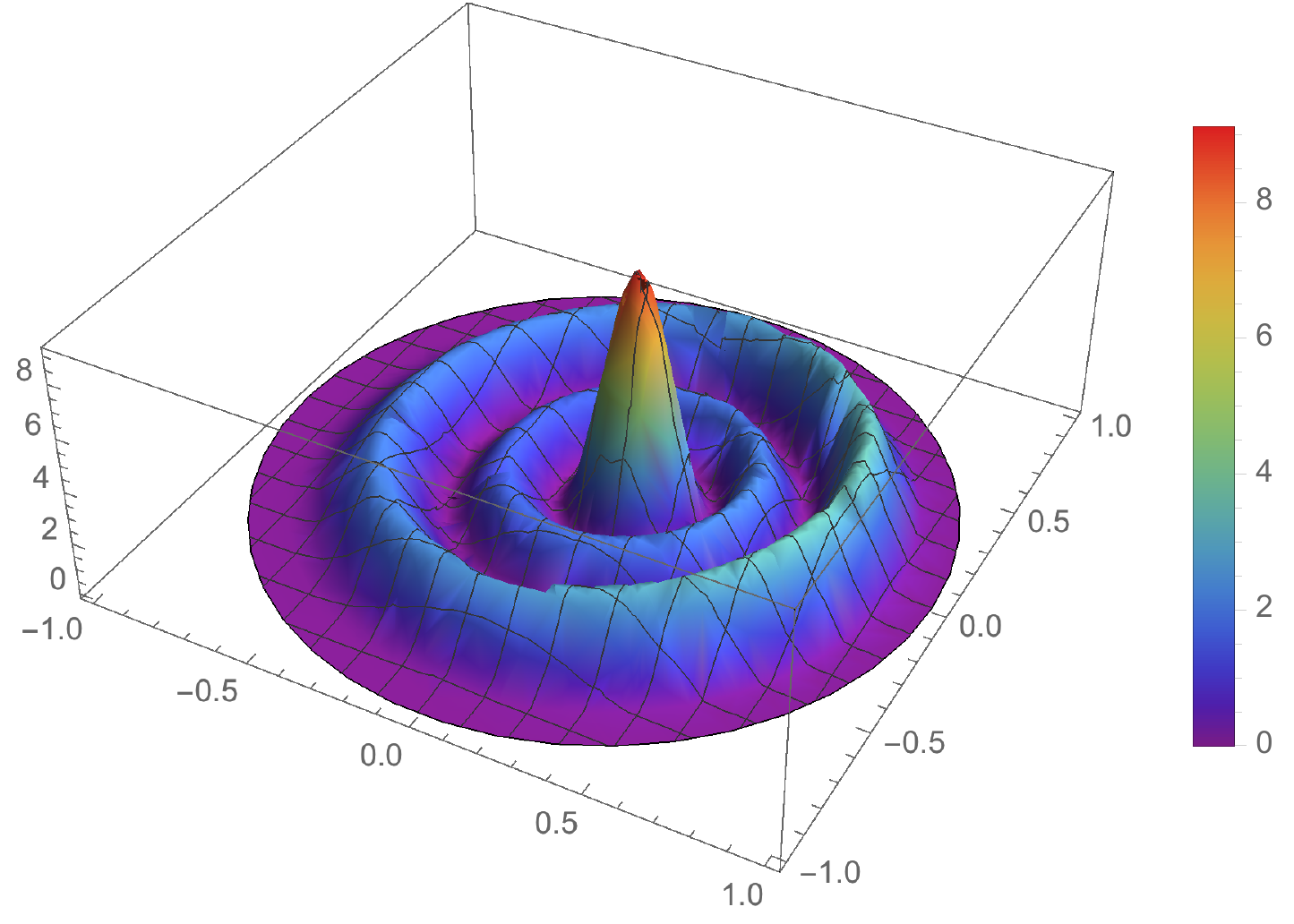} \quad
\includegraphics[width=0.475\textwidth]{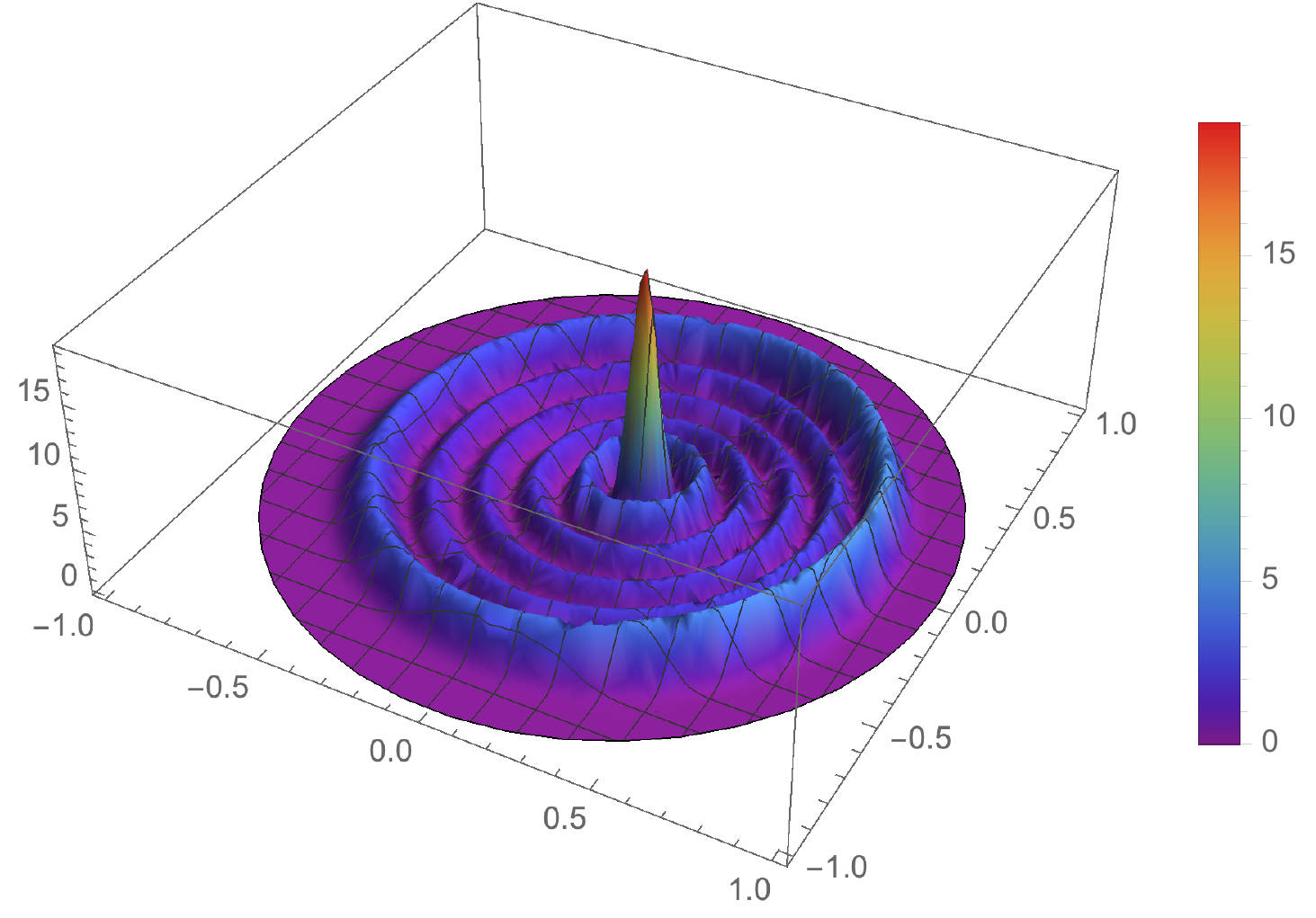}}
\caption{\footnotesize Graphs of $\sfP^{\bsh}_{\alpha}(\beta)$ \eqref{portrait} on the unit disk background for  $\alpha=0.1$,   
  $\varkappa=10$ and different values of $s$.} 
\label{figure9}
\end{figure}

The four graphs displayed in Fig.~\ref{figure8} and Fig.~\ref{figure9} represent the probability distribution 
$\sfP^{\bsh}_{\alpha}(\beta)$  given by \eqref{portrait} on the (purple) unit disk for different values of $s$ and  
$\varkappa$ and $\alpha=0.1$ in all the cases. Thus we consider $\varkappa=5$  and $s=0,1$ in Fig.~\ref{figure8}, and  $\varkappa=10$  and $s=2,5$ in Fig.~\ref{figure9}. Note that  all graphs have a peak centered  in $\alpha=0.1$. The peak height increases when  the number of initial added photons ($s$) increases and the number of  crests surrounding the peak is equal to $s$.
%%%%%%%%%%%%%%%%%%
%%%%%%%%%%%%%%%%%%

\subsection{Non-standard CS quantization of simple functions}

\subsubsection{General}

When applied to the  simplest functions $\alpha$ and $\bar\alpha$, possibly  weighted by a positive $\mathfrak{n}\left(\vert\alpha\vert^2\right)$,   the quantization map \eqref{ANCSquant} yields lowering and raising operators
\begin{equation}\label{lowering}\begin{array}{lllll}
  \mathfrak{n}\left(\vert\alpha\vert^2\right) \alpha & \mapsto   a^{\bsh} &= &\ds\int_{\vert \alpha\vert < R}\frac{\ud^2\alpha}{\pi} \,\tilde w(\vert \alpha\vert^2)\, \alpha\,|\alpha\rg\lg\alpha| \\ 
&   &=&\ds \sum_{n=1}^{\infty}a^{\bsh}_{n-1 n}|n-1\rg\lg n|\, , \\[0.4cm]
 \mathfrak{n}\left(\vert\alpha\vert^2\right) \bar \alpha&\mapsto \left(a^{\bsh}\right)^{\dag} &= &\ds
   \sum_{n=0}^{\infty}\overline{a^{\bsh}_{n n+1}}|n+1\rg\lg n|\, , 
\end{array}\end{equation}
where $\tilde w(u):= \mathfrak{n}(u)w(u)$.
Their matrix elements are given by the integrals
\begin{equation*}
\label{ann}
a^{\bsh}_{n-1 n}=\lg n-1 |a^{\bsh} |n\rg := \int_0^{R^2}\ud u\, \tilde w(u)\,u^n\, h_{n-1}(u)\, \overline{h_n(u)}\, , 
\end{equation*}
and $a^{\bsh}|0\rg = 0$. 

The lower symbol of $a^{\bsh}$ and its adjoint read respectively:
\begin{equation*}
\label{checkaa}
\widecheck{a^{\bsh}} (\alpha)=\lg \alpha| a^{\bsh} |\alpha\rg = \alpha \,\tau\left(\vert \alpha\vert^2\right)\, , \quad \widecheck {\adgh}(\alpha)
= \overline{\widecheck{a^{\bsh}} (\alpha)}\, , 
\end{equation*}
in which the ``weighting''  factor is given by 
\[
\tau(u)= \sum_{n\geq 0} a^{\bsh}_{n n+1}\,u^n\, \overline{h_{n}(u)}\, h_{n+1}(u)\,.
\] 

%%%%%%%%%%%%%%%%%%
%%%%%%%%%%%%%%%%%%
%%%%%%%%%%%%%%%%%%
%%%%%%%%%%%%%%%%%%%%

\subsubsection{The case of SU$(1,1)$ CS}

We consider here the quantisation provided by  the Perelomov SU$(1,1)$ CS  described by \eqref{morecssu11} in Section \ref{SUIICS}. It has been proved in \cite{gazolmo18-1} (Proposition VI.I) the following quantisation formula  
\begin{equation*}
\label{Kpmsu11q}
\frac{2\varkappa-1}{\pi}\int_{\mathcal{D}}\frac{\ud^2 \alpha}{\left(1-\vert\alpha\vert^2\right)^2} \,\frac{2\alpha}{1-\vert\alpha\vert^2} |\alpha; \varkappa;s \rg \lg \alpha; \varkappa;s | = \frac{\varkappa + s}{\varkappa (\varkappa -1)}K_-\,,
\end{equation*}
which is valid for $\varkappa >1$. Hence, the corrective factor to the weight function \eqref{wsu11}  $w(u)= (2\varkappa -1) (1-u)^{-2}$ in order to have $\alpha \mapsto K_-$ in \eqref{lowering} is given by
\begin{equation*}
\label{corrsu11}
\mathfrak{n}(u)= \frac{2\varkappa(\varkappa -1)}{\varkappa +s}(1-u)^{-1}\, . 
\end{equation*}

 \subsection{Non-standard CS from displacement operator}
 
 One can attempt to build (other?) non-standard CS by following the standard procedure involving the unitary ``displacement'' operator built from $a^{\bsh}$ and ${a^{\bsh}}^{\dag}$ 
 \[
 D_{\bsh}(\breve\alpha):= e^{\breve\alpha {a^{\bsh}}^{\dag} -\overline{\breve\alpha} a^{\bsh}}
 \]
 and acting on the vacuum
 \begin{equation}
\label{CDdispv}
|\breve\alpha\rg_{\mathrm{disp}}:= D_{\bsh}(\breve\alpha) \, |0\rg = \sum_{n=0}^{\infty}\breve \alpha^n\,h_n^{\mathrm{disp}}(\vert\breve \alpha\vert^2)\,|n\rg\, , 
\end{equation}
where the notation $\breve\alpha$ is used to make the distinction from the original $\alpha$.  Of course, ${D_{\bsh}}^{\dag}(\breve\alpha)= {D_{\bsh}}^{-1}(\breve\alpha)$ is not equal in general to $D_{\bsh}(-\breve\alpha)$. Besides the example  \eqref{SU11disp} encountered in the  SU$(1,1)$ CS construction, and for which $\tilde\alpha$ is the $\varrho_{\alpha}$ introduced in  \eqref{SU11disp}, and  also  for SU$(2)$ CS \cite{gazeaurev18}, for which the respective weights $\mathfrak{n}(u)$ can be given explicitly, another  recent interesting example  is given in \cite{rodriguezlara17}. In general, it might be quite challenging to determine the functions $h_{n}^{\mathrm{disp}}$ in the expansion \eqref{CDdispv}. 

  More interesting yet is the fact that  these new CS's might show up
in the Glauber's way \cite{glauber63-2}, if we decide to quantize not the simple amplitude of a classical one-mode e.m. field, but its amplitude (i.e. $\alpha$) weighted by a function depending on the intensity of the field (i.e. $\mathfrak{n}(\vert\alpha\vert^2)\alpha$), thus departing  from the historical Dirac (canonical) quantization procedure \cite{dirac27a} to eventually obtain the operators $a^{\bsh}$ and ${a^{\bsh}}^{\dag}$. Hence one introduces a kind of duality between two families of coherent states, the first one used in the quantization procedure $f(\alpha) \mapsto A^{\bsh}_f$, producing the operators $\mathfrak{n}(u)\,\alpha \mapsto a^{\bsh}$ and $\mathfrak{n}(u)\,\bar\alpha \mapsto {a^{\bsh}}^{\dag}$, and so the unitary ``displacement-like''  $D^{\bsh}(\breve\alpha):= e^{\breve\alpha {a^{\bsh}}^{\dag} -\overline{\breve\alpha} a^{\bsh}}$, while the other one uses this $D_{\bsh}(\breve\alpha)$ to build potentially experimental CS yielded in the Glauber's way. 

%%%%%%%%%%%%%%%%%%%%%%%%%%%%%%%%%%%%%%%%%%%%%%%%%%%%%%
%%%%%%%%%%%%%%%%%%%%%%%%%%%%%%%%%%%%%%%%%%%%%%%%%%%%%%

\section{Conclusion}
\label{conclu}
In this work we have presented  families of  coherent states  that are potentially relevant to quantum optics. Of course,  their experimental observation or production comes close to being enigmatic with the current experimental 
physics, at the exception of a few case, like $\varkappa=1/4$ or $\varkappa=3/4$ and $s=0$. Nevertheless, when one considers the way quantum optics has emerged from the golden twenties of quantum mechanics,
nothing prevents us to apply our quantization approach to  classical e.m. fields with weighted intensity. Such classical models might be encountered in non-linear optics. We  think to elementary examples like the time independent or time dependent optical Kerr effect (see for instance \cite{butcher08,boyd2020}). By following the consistent method exposed in the previous section, their quantum counterparts would be  superpositions of new lowering and raising operators like the ones produced by the Perelomov SU$(1,1)$ CS's.

%%%%%%%%%%%%%%%%%%%%%%%%%%%
%%%%%%%%%%%%%%%%%%%%%%%%%%%

\section*{Acknowledgments}  
Acknowledgment. J.P.G. is indebted to the University of Valladolid for its hospitality. Support was provided by MCIN with funding from the European Union NextGenerationEU (PRTRC17.I1) and the PID2020-113406GB-I0 project by MCIN of Spain.

%%%%%%%%%%%%%%%%%%%%%%%%%%%%%%%%%%%%%%%%%%%
%%%%%%%%%%%%%%%%%%%%%%%%%%%%%%%%%%%%%%%%%%%


\begin{thebibliography}{99}

\bibitem{wodeb85} K. Wodkiewicz and J. H. Eberly, Coherent states, squeezed fluctuations, and the SU(2) and SU(1,1) groups in quantum-optics applications, \textit{J. Am. Opt. Soc. B} \textbf{2}  (1985) 458-466.

\bibitem{gerry85} C.~C Gerry, Dynamics of SU(1,1) coherent states,  \textit{Phys. Rev.} \textbf{31} (1985) 2721-2723.

\bibitem{gerry87} C.~C Gerry, Application of SU(1,1) coherent states to the interaction of squeezed light in an anharmonic
oscillator,  \textit{Phys. Rev.} \textbf{35} (1987) 2146-2149.

\bibitem{dellanno06} F. Dell’Anno, S. De Siena, and F. Illuminati,   
Multiphoton quantum optics and quantum state engineering, \textit{Phys. Rep.} \textbf{428} (2006) 53-168. 

\bibitem{gazeaurev18} J.-P. Gazeau, Coherent states in Quantum Optics: An oriented overview, in {\sl Integrability, Supersymmetry and Coherent States}, S. Kuru {\sl et al eds.},  CRM Series in Math. Phys., Springer, Cham (Switzerland)  2019, pp.  69-101.


\bibitem{perel72} A.~M. Perelomov,  Coherent States for Arbitrary Lie Group, \textit{Commun. Math. Phys.} \textbf{26}, (1972) 222-236.  

\bibitem{perel86} A.~M. Perelomov, \textit{Generalized Coherent States and Their Applications} (Springer, Berlin, 1986).

\bibitem{gazeaubook09} J.-P. Gazeau,  
\textit{Coherent States in Quantum Physics}  (Wiley-VCH, Berlin, 2009).

\bibitem{gazolmo18-1} J.-P. Gazeau and M. del Olmo, Covariant integral  quantization of the unit disk, \textit{J. Math. Phys.} \textbf{61}, (2020).  

\bibitem{curadoetal20} E.M.F. Curado, S. Faci, J.-P. Gazeau, D.  Noguera, Lowering Helstrom Bound with non-standard coherent states, \textit{J. Opt. Soc. Am. B} \textbf{38} (2021), 3556-3566. 

\bibitem{glauber63-1} R.~J. Glauber,
          Photons correlations,  \textit{Phys. Rev. Lett.} \textbf{10} 
            (1963) 84-86. 

\bibitem{glauber63-3} R.~J. Glauber, 
          Coherent and incoherent states of radiation field,  \textit{Phys. Rev.} \textbf{131} 
            (1963) 2766-2788.

\bibitem{sudarshan63} E.~C.~G. Sudarshan,  Equivalence of semiclassical and quantum 
		mechanical   descriptions  of statistical light beams,   
		\textit{Phys. Rev. Lett.} \textbf{10} (1963) 277-279.	
		
\bibitem{mandel_wolf70} L. Mandel and E. Wolf,  \textit{Selected Papers on Coherence and Fluctuations of Light}, vols. 1 and 2, Dover, New York, 1970.

\bibitem{mandel_wolf95} L. Mandel and E. Wolf,  {\it Optics Coherence and Quantum Optics} 
(Cambridge University Press, Cambridge, 1995).

\bibitem{iqbal22} S. Javed, H.~B. Monir, N. Amir, and S. Iqbal, Engineering nonclassical SU(1,1) coherent states of light by multiphoton excitation, \emph{Laser Phys.} \textbf{32} (2022) 115201. 

\bibitem{magnus66}    W. Magnus, F. Oberhettinger, and R.~P.  Soni.
\newblock {\em Formulas and Theorems for
the Special Functions of Mathematical Physics}
 \newblock (Springer-Verlag,  Berlin, Heidelberg and New York, 1966).

\bibitem{fox06} M. Fox, \textit{Quantum Optics: An Introduction} (Oxford University Press, New York, 2006).

\bibitem{aharonov73} Y. Aharonov, E.~C. Lerner, H.~W. Huang and J.~M. Knight, Oscillator phase states, thermal equilibrium and group representations, \textit{J. Math. Phys.} \textbf{14} (2011) 746-755.
 
\bibitem{algahel08} S.~T. Ali, J.-P. Gazeau, and B. Heller,  Coherent states and Bayesian duality, \textit{ J. Phys. A: Math. Theor.} \textbf{41} (2008) 365302-1-22. 

\bibitem{stoler70} D. Stoler, Equivalence Classes of Minimum Uncertainty Packets, \textit{Phys. Rev. D} \textbf{1} (1970) 3212-3219.

\bibitem{curadoetal22} E.M.F. Curado, S. Faci, J.-P. Gazeau, D.  Noguera, Helstrom bound for squeezed coherent states in binary communication. \textit{Entropy} \textbf{24} (2022), 220.
 

\bibitem{rodriguezlara17} C. Huerta Alderete, Liliana Villanueva Vergara, and B.~M. Rodr\'{\i}guez-Lara, Nonclassical and semiclassical para-Bose states, \textit{Phys. Rev. A}  \textbf{95} (2017)  043835-1-7. 

\bibitem{glauber63-2} R.~J. Glauber, 
          The quantum theory of optical coherence,  \textit{Phys. Rev.} \textbf{130} 
            (\textbf{1963}) 2529-2539. 

\bibitem{dirac27a} P.~A.~M. Dirac, The Quantum Theory of Emission and Absorption of Radiation, \textit{Proc. Royal Soc. Lond. A} \textbf{114} (1927) 243-265.  

\bibitem{lvovsky20} I. Lvovsky, P. Grangier, A. Ourjoumtsev, V. Parigi, M. Sasaki, and R. Tualle-Brouri, Production and applications of non-Gaussian quantum states of light,  arXiv:2006.16985v1 [quant-ph] 

 \bibitem{faghihi20} M.~J. Faghihi,  Generalized Photon Added and Subtracted $f$-Deformed Displaced Fock States,\textit{ Ann. Phys.} (Berlin) \textbf{532} (2020) 2000215.
 
\bibitem{swain22} S. N. Swain,Y. Jha, and  P.~K. Panigrahi,   Two-mode photon added Schr\"{o}dinger cat states: nonclassicality and entanglement,  \textit{J. Opt. Soc. Am. B} \textbf{39}  (2022) 558-565.

\bibitem{butcher08} P.~N. Butcher and D. Cotter, \textit{Elements of Nonlinear Optics} (Cambridge University Press, 2008).

\bibitem{boyd2020} R.~W. Boyd, \textit{Nonlinear Optics} (Amsterdam,  Academic Press, 2020).

\end{thebibliography}
\end{document}